\documentclass{article}

\usepackage[a4paper,margin=3cm]{geometry}
\usepackage{setspace}
\usepackage{natbib}
\usepackage[none]{hyphenat}
\usepackage{float}
\usepackage{multirow}
\usepackage[compact]{titlesec}
\usepackage{amsmath, bm}
\usepackage{amssymb}
\usepackage{authblk}
\usepackage{enumitem}
\usepackage{xfrac}
\usepackage{booktabs}
\usepackage{ragged2e}
\usepackage{csquotes}
\usepackage{pdflscape}
\usepackage[title]{appendix}
\usepackage{xcolor}
\usepackage{hyperref}
\usepackage{graphicx}
\usepackage{adjustbox}

\hypersetup{
    colorlinks,
    linkcolor={blue!80!black},
    citecolor={blue!80!black},
    urlcolor={blue!80!black}
}

\title{The Complexity of Corporate Culture as a Potential Source of Firm Profit Differentials}
\author[a]{Frederik Banning\thanks{Corresponding author. \href{mailto:frederik.banning@ruhr-uni-bochum.de}{frederik.banning@ruhr-uni-bochum.de}}}
\author[a]{Jessica Reale}
\author[a]{Michael Roos}
\affil[a]{Chair of Macroeconomics, Ruhr University Bochum, Universitätsstraße 150, 44801 Bochum, Germany\thanks{Funding: This work was supported by Deutsche Forschungsgemeinschaft (DFG) [426603947 granted to MR]. The funder had no role in study design, data collection and analysis, decision to publish, or preparation of the manuscript.}}

\date{}

\begin{document}

\maketitle

\begin{abstract}
\justifying \sloppy
    This paper proposes an addition to the firm-based perspective on intra-industry profitability differentials by modelling a business organisation as a complex adaptive system.
    The presented agent-based model introduces an endogenous similarity-based social network and employees' reactions to dynamic management strategies informed by key company benchmarks.
    The value-based decision-making of employees shapes the behaviour of others through their perception of social norms from which a corporate culture emerges.
    These elements induce intertwined feedback mechanisms which lead to unforeseen profitability outcomes.
    The simulations reveal that variants of extreme adaptation of management style yield higher profitability in the long run than the more moderate alternatives.
    Furthermore, we observe convergence towards a dominant management strategy with low intensity in monitoring efforts as well as high monetary incentivisation of cooperative behaviour.
    The results suggest that measures increasing the connectedness of the workforce across all four value groups might be advisable to escape potential lock-in situation and thus raise profitability.
    A further positive impact on profitability can be achieved through knowledge about the distribution of personal values among a firm's employees.
    Choosing appropriate and enabling management strategies, and sticking to them in the long run, can support the realisation of the inherent self-organisational capacities of the workforce, ultimately leading to higher profitability through cultural stability.
\end{abstract}
% american english is recommended for keywords
\bigskip
% JEL => C63: Simulation modeling; D21: Firm Behaviour: Theory; L25: Industrial Organization: Firms' profits; M14: Corporate Culture; Z13: Social norms and social networks

\newpage
% ----- begin main body
\section{Introduction}
\justifying \sloppy
% problem/motivation
Firms operating within the same industry have long experienced persistent profit margin differentials \citep[see][among others]{Mueller.1986, Geroski.1988, Leiponen.2000}.
Intra-industry profitability differentials sparked a long-standing debate about which factors drive firms' performance \citep[for a thorough review of productivity determinants see][]{Syverson.2011}.
The dichotomy between industry- and firm-based factors has led to two competing theories \citep{Bowman.1990}.
The industrial organisation (IO) literature addresses profit differentials to structural characteristics of the firms' industry.
Assuming homogeneous firms \citep{Mauri.1998}, IO-based empirical research has tied higher profits to (i) market power and concentration ratios \citep{Bain.1954}, (ii) entry barriers \citep{Mann.1966}, (iii) mobility barriers \citep{Semmler.1984}\footnote{
Mobility barriers comprise factors hindering a firm's entry and exit from an industry.
}, and (iv) market shares \citep{Ravenscraft.1983}.

The resource-based view (RBV) of the firm challenges the implicit assumption of firms' homogeneity behind the industry-driven empirical analyses.
The RBV emphasises firm-specific factors as the drivers of performance variance between firms within the same industry \citep{Hansen.1989, Cefis.2005, Blease.2010}.
In this vein, each organisation has a \textit{unique} set of resources and capabilities, which include management skills \citep{Bowman.2001}, routines \citep{Barney.2001} and corporate culture \citep{Barney.1986, Gorton.2022, Grennan.2023}.
Existing literature has provided stronger empirical support for firm effects as drivers of performance and profit differentials \citep[see][among others]{Hawawini.2004, Arend.2009, Bamiatzi.2009, Hambrick.2014, Bamiatzi.2016, Backus.2020}.
Firms' idiosyncratic resources have thus greater explanatory power for competitive advantage than industry-specific effects \citep{Galbreath.2008}.
Moreover, the prevalence of industry effects on firm factors \citep{Schmalensee.1985} has been observed less frequently and is valid for medium-size firms only \citep{Fernandez.2019}.

The economic literature often treats firms as a \enquote{black box}.
As such, organisations are conceived as indivisible units with homogeneous resources and capabilities where firm-specific factors have no relevance.
In order to better understand the sources of competitive advantage and profit differentials, opening up the black box is necessary to avoid neglecting the complexity of firms' internal structure \citep{Foss.1994}.
Moreover, existing literature has mainly focused on the response of firms' resources to \enquote{shifts in the business environment}\citep[][page 515]{Teece.1997}.
However, it is still unclear how company-specific factors, i.e. corporate culture and management strategies, co-evolve and impact corporate performance when firms' inner conditions change.

% paper content and research questions
This article contributes to the understanding of the inner workings of a firm by (i) combining concepts and methods from economics, (social) psychology, and complexity science, and (ii) adopting an agent-based bottom-up approach.
We regard formal and informal institutions within a firm as a key factor of independent and heterogeneous human resources embedded in a social context, interacting actively and reactively with other employees and responding to corporate strategy changes.
To capture these mechanisms, we see a company as an example of a complex adaptive system (CAS) \citep{Fuller.2001}, inherently difficult to control and manage by nature \citep{Holland.1992, Fredin.2020}.

This paper extends an earlier agent-based model \citep{Roos.2022}.
It builds on top of its framework within which agents have heterogeneous value hierarchies \citep{Schwartz.2012b}, shaping corporate culture – captured via descriptive social norms \citep{Deutsch.1955}\footnote{
    Descriptive social norms refer to what is \textit{seen as normal} within an organisation, formalised in \cite{Roos.2022} as the average behaviour of all employees.
}
– and mediating the impact of management instruments on corporate performance.\footnote{
    \cite{Foster.2000} analyses the relation between the transaction-cost framework and CAS-based views of business organisations.
    Specifically, the usefulness of the Coasian tradition for complex adaptive approaches is seen through the lens of the \textit{psychological complexity} embedded in the formalisation of the opportunism-oriented \enquote{contractual man}, which goes beyond the usual rational homo oeconomicus on which the neoclassical tradition is grounded.
    While we deem this interpretation worth mentioning, we still feel the urge to complement a CAS-firm approach with multiple and heterogeneous employees' motivational dispositions – i.e. only accounting for opportunism is not sufficient – by modelling individual decision-making based on social norms and personal values in a \textit{self-contained} corporate environment. 
    This CAS-based extension thus does not aim at a theoretical reconciliation of previous traditions with complexity science.
    On the contrary, we want this paper to provide a new, modular, and extensible framework to serve as an openly accessible foundation for future research on modelling firms as complex adaptive systems.
}
As first steps towards modelling an organisation as a CAS, we propose three main extensions.
First, agents take part in an endogenous and dynamic social network, whose peering mechanism determines how social norms and corporate culture spread within the organisation. 
Second, employees heterogeneously adapt their behaviour to corporate strategies, depending on how management instruments – i.e., monitoring and monetary incentives – affect their value-based satisfaction.
Third, the management can more or less frequently update the degree of its monitoring activities and the implementation of pay-for-performance (PFP) schemes, which are influenced by, and feed back into, the development of corporate culture.
By combining a dynamic corporate culture, employees' adaptive behaviour, and endogenous management strategies, we aim to answer the following research questions:
What are potential effects of corporate culture on the profit differentials of otherwise similar firms?
In which ways are corporate outcome and profitability affected by the frequency of changes in management strategy?
Under which conditions – if any – will the management's attempts to steer the organisation boost profitability?

% relevance
Integration and management of firms' resources are essential for productivity \citep{Russo.1997}.
Organisations depend on the management's efforts to reach – potentially incompatible –  corporate goals like maximising profits, output, employee cooperation and satisfaction. In such a context, the instruments available to organisational management can yield unintended or unexpected dynamics that are not easily foreseeable and might derive from 
    (i) the interaction between different management instruments and
    (ii) agents' heterogeneous response to the implemented corporate strategies, leading to adaptive behavioural patterns.
Firms are also organisations that form a \enquote{minisociety} itself \citep{Macneil.1977}, whose members struggle for an outcome that considers all the values, attitudes, ideas, history, resources, and skills of relevant individuals, in-groups and out-groups, and society as a whole.\footnote{
    Business scholars have long identified this phenomenon under the concept of \textit{stakeholder management}, which has led to a growing interest in stakeholder theory as a rival paradigm to the contractual perspective \citep{Key.1999}.
    The management literature has been involved in a long-lasting debate about similarities and divergences between the competence- and stakeholder-based theories of the firm \citep[see][among others]{Hodgson.1998, Freeman.2021}.
    Since this controversy lies beyond the scope of this paper, we conceive the two approaches as non-rivals in light of (i) their common origination in the strategic management field, (ii) their mutual influence, and (iii) their rejection by the dominant paradigm.
}
Neglecting the intertwined nature of managerial boundaries and the action space of the affected agents might lead to ineffective management and, thus, to undesirable outcomes for the entire system. 
Theories of the firm that ignore the social environment of organisations might also encourage a \enquote{practically unsustainable neglect of society by actual firms} \citep[page 1062]{Thompson.2017} and might thus produce bad public policies \citep{Teece.2017}.

% Gap in the literature
This paper fills a gap in the economic literature by adopting a multidisciplinary approach to the theory of the firm, combined with an unambiguous decolonisation of the field from the maximising modes of agents’ behaviour.
In an attempt to ”’open-up’ the black box” \citep{Casson.2005}, contract-based theories of the firm in the Coase tradition focused on employers’ and employees’ decision-making as driven by internal and external transaction costs.
However, notions of profit maximisation and market equilibrium – at the core of this transaction-cost framework – make behavioural changes merely dependent on extrinsic causes.
By identifying agents’ competencies as the cornerstone of the theory of the firm, the evolutionary (competence) approach emphasises the crucial role of the social component within organisations \citep{Foss.1993}.
While acting as a foundation for the monolith of literature in economics and business that deals with firms as such ultra-optimised institutions, the underlying assumptions of the competence-based approach – highly rational and fully informed agents who maximise their expected utility – still advance an inadequate representation of firms’ behaviour as a “series of rational and dispassionate activities” \citep[][page 1501]{Hodgkinson.2011}.
As a consequence, standard microfoundations – also in the strategic management tradition – appear considerably incompatible with the findings of experimental economists, psychologists, sociologists, neuroscientists, and others (e.g. \citealp{Kahnemann.1979, Shafir.2002, Sarnecki.2007}).
Moreover, the two predominant theories of the firm – contract-based and competence-based – neglect social behaviour.
Contrary to empirical findings, they thus overlook (i) the \textit{glue role} of corporate culture \citep{Freiling.2010, Heine.2013} in binding tangible, intangible, and personnel-based resources – proper of the RBV \citep{Grant.1991} – and (ii) the impact of firm-specific factors on profitability.

Even though building models of complex systems is necessarily a reductionist approach towards real-world practices and processes, it allows us to study firms' dynamics within predefined limits – inherent to the CAS approach – and to assume more flexible behavioural rules suitable for several modelling scenarios.
Moreover, conceptualising a firm as a CAS within confined boundaries and spheres of influence facilitates the study of its internal workings and allows us to use it as a laboratory for systems with higher complexity \citep{Guiso.2015}.
%4. What have other people done that is related to the paper?
There is a relatively young tradition of scholars working on firms as CAS.
The focuses of these previous studies have been widespread and range from innovation \citep{Chiva.2004, Akgun.2014, Inigo.2016}, to entrepreneurial ecosystems \citep{Roundy.2018, Fredin.2020}, knowledge diffusion \citep{Magnanini.2021}, and learning \citep{Marsick.2017, Lizier.2022}. 
This paper contributes to this line of thought both in terms of object of study – i.e. business organisations – and employed method – CAS-based approach – but focuses, distinct from already existing literature, on corporate culture and its influence on firm performance and profitability.

The paper is structured as follows.
Section \ref{sec: model} describes the three extensions of the model, i.e. network formation and the emergence of corporate culture, employees' adaptive behavioural rules, and endogenous management strategies.
Section \ref{sec: results} explains the simulations and the main results of the model, and section \ref{sec: discussion} discusses the relevance of our findings. 
The last section concludes.

\section{Model}\label{sec: model}
There are $n$ employees in the company.
Every employee $i \in N$ where $N = [1,n]$ has the same daily time budget $\tau$, which has to be allocated among three activities: cooperation ($c_{i,t}$), shirking ($s_{i,t}$) and, residually, individual tasks ($p_{i,t}$).
Employees' behaviour depends on personal values \citep{Schwartz.2012b}.
Each agent belongs to one of the four higher-order value types:
Self-transcendent (ST-type) employees are motivated by benevolence and universalism, and self-enhancing (SE-type) agents by power and achievement.
Conservative (C-type) employees value security and conformity above all else, whereas open-to-change individuals (O-type) especially value self-direction and stimulation.
Agents' decisions depend on social norms, from which they can deviate positively or negatively \citep{Li.2021a}.
Time allocations among the three activities are assumed to be triangularly distributed and are modelled in terms of stochastic deviations from the cooperative ($c^{*}_{i,t}$) and shirking ($s^{*}_{i,t}$) norms, defined in section \ref{subsec: network}. 
The main behavioural equations which constitute the backbone of this model follow \cite{Roos.2022}.
For the sake of clarity, Table \ref{tab: list_eq} in Appendix \ref{sec: app-1} provides a brief comparison between the original equations and the changes the three additional extensions presented in this paper entail.

The management can implement monitoring strategies and/or financial rewards. 
The adoption of these instruments can lead to a certain degree of trusting (or controlling) management style and/or to a competitive or cooperative rewards setting (PFP schemes).\footnote{
    A full list of starting values for the model parameters can be found in Table \ref{tab: parameters} of Appendix \ref{sec: app-1}.
}

\begin{figure}[t]
    \centering
    \includegraphics[scale = 0.7]{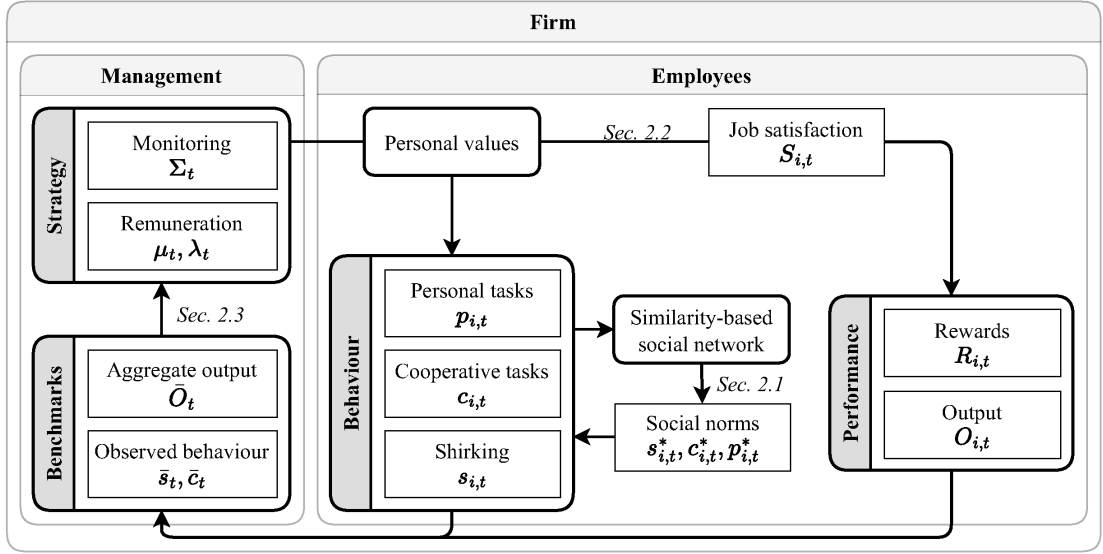}
    \caption{Model overview. Source: Authors' own illustration.}
    \label{fig: model-overview}
\end{figure}

In the following subsections, we explain in detail (i) the network formation and its influence on social norms and corporate culture (\ref{subsec: network}), (ii) employees' adaptive behaviour based on job satisfaction concerns (\ref{subsec: adaptive-behaviour}), and (iii) how management strategies endogenously react to key company benchmarks (\ref{subsec: end_management}).
For reasons of clarity and comprehensibility, Figure \ref{fig: model-overview} presents the model overview, highlights the main feedback mechanisms our CAS-based firm is based on (bold lines), and indicates for each extension its corresponding section.  \bigskip

\subsection{Spread of norms in a social network}\label{subsec: network}
In \citet{Roos.2022}, employees' perceived social norms for each task are assumed to be equal to the actual descriptive norm inside the firm, modelled as the mean behaviour among all agents.
In this paper, we relax this artificial assumption of static environments by modelling the spread of information about social norms via the construction of an endogenous network where informal connections in the firm (e.g. based on value homophily or cooperation and shirking intensity) continuously evolve over time.
These connections form a personal network that captures an employee's relevant peers and, thus, also their perceived social norms.
Since all the intricate details of personal interactions within a firm – no matter if directly related to work or non-work processes – are intractable, we need to provide a formalised simplification able to capture the inherent dynamics of the endogenous formation and evolution of such a social network. For this purpose, we exploit the tenets of the Social Referent Literature (SRL).

The SRL distinguishes two types of social referents at the workplace \citep{Brass.1984}. On the one hand, \textit{cohesive} referents are the ones with whom employees engage more frequently, potentially allowing for the formation of close interpersonal ties and friendships \citep{Galaskiewicz.1989}. On the other hand, \textit{structurally equivalent} relationships are formed among employees who perform the same role, occupy the same position in a network, or share a similar pattern of relationships \citep{Burt.1987}.
There is a long-standing debate about what kind of information the two kinds of actors (referents) are likely to share with co-workers which has revealed interesting findings.
Cohesive referents share more general organisational information related to employees' social integration within the firm's corporate culture, while employees turn to structural referents for information strictly related to their work – such as tasks, roles and responsibilities – for performance improvements \citep{Shah.2000}.
While not neglecting the influence of structural relationships on employees' social networks, we currently leave aside considerations about formal connections.
Indeed, in the network we aim to construct, co-workers' interactions inform employees about the prevailing social norm within the firm, information less likely shared by structural referents \citep{Shah.1998}.
Therefore, behavioural norms – i.e. normative information\footnote{
    \cite{Shah.1998} classifies normative information under the category of \enquote{general organizational information}, relevant for social adaptation to a company's culture and social integration within a group.
    While the author specifically refers to \enquote{norms of expected behaviour} – i.e. injunctive norms, which are outside the scope of this paper – we deem descriptive norms as also falling in the above-mentioned category as these might encourage behavioural conformity \citep{Cialdini.1998} for the sake of belonging to a social group and being integrated within an organisation's social system, even independently of self-identification concerns \citep{Pryor.2019}.
} – are best acquired via cohesive social relationships, whose formation depends on the \textit{frequency} and the \textit{intensity} of interactions within organisations. 

By exploiting the concept of cohesive relationships as dependent on the frequency of contacts, we propose a \textit{similarity-based} approach that relies on the agents' decisions regarding their time allocation during each workday.
In other words, employees' chances of connection are determined by how similarly they spend their time on the three activities: cooperation ($c_{i,t}$), shirking ($s_{i,t}$) and individual tasks ($p_{i,t}$).
The rationale behind this is that employees have a higher chance to meet others who spend their time in similar ways, resulting in a higher probability to connect with one another while performing these activities:
The higher the activities similarity, the greater the probability of contact between two agents and the higher the amount of transmitted cues about socially normative information.

Assume that the initial connection strength (i.e. edge weight) – how well agents know each other – between two employees $\{i,j\} \in N$ is zero, such that $e_{i,j,t-1} = e_{j,i,t-1} = 0$. 
This means that the simulation starts with a \enquote{blank slate} social network without any connections and fully isolated agents, i.e. with $n$ vertices and $0$ edges.
The set of an agent's peers $P_{i,t} = \{ j \; | \; e_{i,j,t-1} \neq 0 \}$ is therefore empty.
As a consequence, there is no influence of individual behaviour on other agents' perceived social norms during the first time period.
During each subsequent workday, agents make their regular decisions on how to spend their time.
Based on employees' decisions, we calculate the activity differences ($AD_{i,j,t}$) between each pair of agents regarding the time they spent on the three activity types $c_{i,t}$, $s_{i,t}$ and $p_{i,t}$.
This calculation is performed from the perspective of each agent $i$ regarding all other agents and vice versa.

To model this, we exploit and adapt the weighted absolute-deviation index (WADI) proposed by \cite{Stewart.2006}, which facilitates interpretation and guarantees a higher degree of robustness compared to other dissimilarity indices.
Translated into a case-independent equation that is able to deal with any number $A$ of generic activities $a$, this leads us to formalise $AD_{i,j,t}$ as in Equation \ref{eq: WAD-index}.\footnote{
    In this model, performing individual activities ($p_{i,t}$) also enters the calculation of the activity similarity measure.
    Doing so allows modelling of the fact that employees may gain behavioural information from co-workers sharing the same office or working space, even while devoting time to individual tasks.
} 

\begin{equation}\label{eq: WAD-index}
    AD_{i,j,t} = \frac{\sum_{k=1}^{A} \lvert a_{k,i,t} - a_{k,j,t} \rvert}{\tau}
\end{equation}

Equation \ref{eq: WAD-index} describes the weighted absolute difference in activities between two agents with equal daily working time $\tau_{i,t} = \tau_{j,t} = \tau$.
The weights are thus equal to the fraction of the total time spent in each activity\footnote{
    The weights in Equation \ref{eq: WAD-index} are implicit as they are equal to $\sum_{k=1}^{A} \frac{(a_{k,i,t} + a_{k,j,t})}{(a_{k,i,t} + a_{k,j,t})} = 1$, if employees $i$ and $j$ are endowed with the same amount of available maximum working time.
}
and it follows that the activity differences between two employees would be symmetrical, that is $AD_{i,j,t} = AD_{j,i,t}$.\footnote{
    However, interacting pairs of employees may also experience heterogeneous degrees of relational \textit{intensity}.
    When agents have different time budgets ($\tau_{i,t} \; \neq \; \tau_{j,t}$), the spread of social norms might be asymmetrical – i.e. $e_{i,j,t} \; \neq \; e_{j,i,t}$ – reflecting the potential asymmetrical reciprocity of relational ties which exerts a great impact on firms' social dynamics and corporate performance \citep{Lopez.2018}.
    Therefore, each employee might value the three activities differently. In the context of this model, the importance each agent assigns to an activity can be deduced by the relative time spent on that task with respect to its individual time budget. 
    To account for differences in \textit{relative importance} of a given activity for any agent, we could extend the calculation of the WADI by introducing a simple weight $w_{i,t} = \frac{\tau_{i,t}}{\tau_{j,t}}$. However, we leave heterogeneous time budgets to future works dealing with flexible working time arrangements.
}

To calculate employees' activity similarity $AS_{i,j,t}$, we subtract the previously computed activity difference from 1, to reflect the positive impact of perceived activity similarity on agents' connection strength.
The higher the activity similarity, the stronger the employees' connection during this working day.

\begin{equation}
    AS_{i,j,t} = 1 - AD_{i,j,t}
\end{equation}

This similarity measure $AS_{i,j,t}$ is used to represent (i) the chance that the two agents have met during a workday $t$, and (ii) agent $i$'s assigned importance of occurred interactions with employee $j$.
To model the chance of agents' interactions, we make a random draw $d_{i,j,t}$ from a uniform distribution between $0$ and $1$ for each agent pairing.
Let $d_{t}$ denote the set of these draws.\footnote{
    Note that the fixed set $d_{t}$ necessarily means that interactions are always symmetrical between two agents, implicitly assuming that $i$ cannot interact with $j$ without $j$ also interacting with $i$.
}

\begin{equation}\tag{b}
    d_{t} = \{ d_{i,j,t} \sim U(0, 1), \; \forall \; (i,j) \in N \}
\end{equation}

If the value of $d_{i,j,t}$ is less than their activity similarity $AS_{i,j,t}$ ($= AS_{j,i,t}$), they interact during the current workday.

If and how well employees know each other ($e_{i,j,t}$) determines the order by which agent $i$ checks for potential interactions ($I_{i,j}^{pot}$).
Agents will always first check for potential interactions with their existing peers $P_{i,t-1}$, starting from those with whom they have the strongest connection  (i.e. $\max (e_{i,j,t-1}), \; \forall \; j \in P_{i,t-1}$) and going through this sequence in descending order.
Only after that has been done agents also check for potential interactions with other randomly chosen employees who are yet unknown to them.
Each agent $j$ can only be checked once for possible interaction with $i$ and can also only be interacted with once.
\begin{equation}\label{eq: set-potential-interactions}
    I_{i,t}^{pot} = 
      \{j \;|\; e_{i,j,t-1} > e_{i,k,t-1}, \forall \; j,k \in P_{i,t-1},\; j\neq k \} \cup 
       \{j \; | \; j \in_{R} N \setminus P_{i,t-1} \} 
\end{equation}

Therefore, the set of agents with whom employee $i$ interacts ($I_{i,t}$) can be defined following Equation \ref{eq: set-agents-interactions}.

\begin{equation}\label{eq: set-agents-interactions}
    I_{i,t} = \{ j \; | \; d_{i,j,t} \; < \; AS_{i,j,t}, \; \forall  j \in I_{i,t}^{pot} \}
\end{equation}

Equation \ref{eq: set-agents-interactions} stochastically determines whether or not two agents interact, and Equation \ref{eq: set-potential-interactions} captures the order in which potential interactions are checked.
Naturally, this leads to relatively dense networks over time, which is especially evident in the long run if agents' behaviours converge.
Such high amounts of daily interactions stand in stark contrast to empirical findings from epidemiology, which show that on an average daily basis, people have 8 \citep{Leung.2017} or 13.4 contacts \citep{Mossong.2008}.
To avoid the peculiarity of extremely high interactivity in our theoretical model of a firm, a new agent variable is added, which limits the amount of interactions agents can have over the course of one day.
At each step, agents pick their maximum amount of interactions ($\iota_{i,t}$) from a theoretical distribution ($ID$) such that $\iota_{i,t} \sim ID$.\footnote{
    To account for the fact that there are no fractional interactions in our model, $\iota_{i,t}$ is rounded to its nearest integer value.
}
This distribution can either be informed by empirical literature or created freely to explore its effects on the modelling results.
For the analysis conducted throughout this paper, we have chosen a uniform distribution $ID = U(0, 7.14)$ loosely based on the contact numbers of the above-mentioned studies.\footnote{
    Attributing the same weight to their results, we assumed that people have $(8 + 13.4) / 2 = 10.7$ contacts per day on average.
    Further assuming an equal distribution of contacts across the day, we estimate the average amount of work contacts on a normal work day with $\tau = 8$ to be $\frac{8}{24} * 10.7 \approx 3.57$.
}
Therefore, $I_{i,t}$ can never contain more than $\iota_{i,t}$ elements, and, as such, the set is truncated after $\iota_{i,t}$ elements.

Whenever agents $i$ and $j$ interact, their connection intensifies by $AS_{i,j,t}$.
Otherwise, their connection strength does not change during this workday.
We introduce $\Delta e_{i,j,t}$ to reflect the weight change for each agent $i$'s directed edge toward agent $j$.

\begin{equation}
    \Delta e_{i,j,t} =
    \left\{ \begin{array}{ll}
        AS_{i,j,t} & \text{if} \; j \in I_{i,t}\\
        0          & \text{otherwise}
    \end{array} \right. \\
\end{equation}

It describes how strong the interaction between agents $i$ and $j$ is during that day, and by that also how important agent $j$'s behaviour is for agent $i$'s updating of descriptive social norms.
At the end of the workday $t$, the new connection between agents $i$ and $j$ can be formulated as in Equation \ref{eq: edge_weights}.

\begin{equation}\label{eq: edge_weights}
    e_{i,j,t} = 
    \left\{ \begin{array}{ll}
        \frac{(t-1) \cdot e_{i,j,t-1} + \Delta e_{i,j,t}}{t} & \text{if}\; t \geq 1 \\
        0                                                    & \text{if}\; t = 0
    \end{array} \right. \\
\end{equation}

The edge weights $e_{i,j,t}$ reflect the long-term interaction history between two agents while also accounting for the fact that the connection between them deteriorates over time if no interaction takes place.
They are then used to update the \textit{descriptive} social norms perceived by agent $i$, describing the relative influence of peers with whom $i$ has interacted during this workday.
This leads to the following adaptations to equations (3) and (4) from \cite{Roos.2022}:

\begin{align}
    \label{eq: shirking-norm}
    s_{i,t}^{*} & = 
    \left\{ \begin{array}{ll}
        (1 - h) \; s_{i,t-1}^{*} + h \; \frac{\sum_{j \in I_{i,t}} \Delta e_{i,j,t-1} \; s_{j,t-1}}{\sum_{j \in I_{i,t}} \Delta e_{i,j,t-1}} & \text{if} \; I_{i,t} \neq \emptyset\\
        s_{i,t-1}^{*}      & \text{otherwise}
    \end{array} \right. \\
    \label{eq: cooperation-norm}
    c_{i,t}^{*} & = 
    \left\{ \begin{array}{ll}
        (1 - h) \; c_{i,t-1}^{*} + h \; \frac{\sum_{j \in I_{i,t}} \Delta e_{i,j,t-1} \; c_{j,t-1}}{\sum_{j \in I_{i,t}} \Delta e_{i,j,t-1}} & \text{if} \; I_{i,t} \neq \emptyset\\
        c_{i,t-1}^{*}      & \text{otherwise}
    \end{array} \right.
\end{align}

Rather than modelling employees' motivation to maintain specific ties at the workplace \citep[see e.g.][]{Randel.2007}, we assume that agents remember all past interactions with others, no matter how weak the ties between them.
Because the strength of each connection can only grow by a value between $[0,1]$ per simulation step, we can observe the \textit{relative} strength (weakness) of emergent connections between agents who have historically interacted more (less) frequently.

\subsection{Adaptive behaviour and job satisfaction}\label{subsec: adaptive-behaviour}

A moderate but robust correlation regarding the effects of job satisfaction on job performance has been found by meta-studies \citep{Judge.2001,Fisher.2003}.\footnote{
    It is noteworthy that while we modelled a direct connection between productivity effects and job satisfaction, there are other plausible relationships dealing with the broad spectrum of employee happiness \citep{Thompson.2021}, organisational citizenship behaviour \citep{Spector.2022}, and counterproductive work behaviour \citep{Nemteanu.2021}.
}
To incorporate this in the model, each employee has a level of job satisfaction $S_{i,t} \in [0,1]$ that directly influences job performance through short-run productivity effects $\pi_{i,t}$.

We assume that $\pi_{i,t} = (1 - S^{eff}) + 2 \cdot S^{eff} \cdot S_{i,t}$ where $S^{eff} \in [0,1]$ is an exogenous model parameter mediating the effect of satisfaction on productivity.\footnote{
    This simplified approach is chosen on a forward-looking basis to facilitate model calibration and later integration of other productivity factors.
}
For the simulations conducted in this paper, we have chosen $S^{eff} = 0.5$, which results in $\pi_{i,t} \in [0.5,1.5]$.
Under these conditions, dissatisfaction (low $S_{i,t}$) directly leads to a reduction in productivity by impacting the intensity with which working time is used and thus individual output ($O_{i,t}$). 

\begin{equation}\label{eq: output_eq}
    O_{i,t} = \pi_{i,t} (p_{i,t}^{\phantom{i}(1-\kappa)}\cdot \bar{c}_{i,t}^{\phantom{i}\kappa})w
\end{equation}

\noindent Individual output thus depends on (i) the time devoted to individual tasks ($p_{i,t}$), (ii) the average cooperative time ($\bar{c_{i,t}} = \sfrac{1}{(n-1)} \sum_{j \neq i} c_{j,t}$), and (iii) the extent to which employee $i$'s performance depends on the support of co-workers ($\kappa$) and on initial productivity.

The firm's management employs a certain degree of monitoring $\Sigma$ which can range between a fully trusting ($\Sigma = 0$) and a fully controlling ($\Sigma = 1$) management style.\footnote{
    In a controlling environment, C-type employees are happiest and shirk much less than the social norm, and the opposite occurs with O-type employees.
    Vice versa under a trusting management attitude.
    SE and ST employees are assumed to be indifferent to monitoring but responsive to financial rewards.
}
The bonus each employee $i$ receives is defined in Equation \ref{bonus}
and depends on the type of PFP scheme ($\lambda$) implemented and individual output $O_{i,t}$.
Pure bonus systems can incentivise only one type of behaviour, i.e. by linking bonus payments to individual ($\lambda = 0$) or joint ($\lambda = 1$) output.
Mixed PFP schemes ($\lambda = [0,1]$) also cover intermediate cases where a proportional combination of output assessment is used.

\begin{equation}\label{bonus}
    B_{i,t} = (1-\lambda)O_{i,t} + \lambda (\frac{1}{n}) \sum_{j=1}^{n} O_{j,t}
\end{equation}

The firm pays all employees a homogeneous base wage $\omega_{b}$, plus individual bonuses ($B_{i,t}$), which are weighted for a parameter $\mu = \{0,1\}$ that reflects the intensity of rewards the management is willing to offer.\footnote{
    The intensity of PFP schemes $\mu$ is such that $\mu = 0$ if no rewards are implemented, and $\mu = 1$ if bonuses are granted, whatever the type.
}

\begin{equation}\label{rewards}
    R_{i,t} = \omega_{b} + \mu B_{i,t}
\end{equation}

An employee's base satisfaction level ($S_{i}^{0}$) can take a value between $[0,1]$ where $0$ is completely dissatisfied and $1$ means completely satisfied, therefore defining a neutral level of job satisfaction to be at $0.5$.
Equation \ref{eq: base-satisfaction} shows that it is dependent on the employees' value types, management's monitoring efforts ($\Sigma$), the type of implemented PFP schemes ($\lambda$), and their intensity ($\mu$).
The initial satisfaction of each agent at the beginning of the simulation is equal to $S_{i,t=0} = S_{i}^{0}$.

\begin{equation}\label{eq: base-satisfaction}
    S_{i}^{0} = 
    \left\{ \begin{array}{ll}
        \Sigma                    & \text{if}\; i\;\in\; \text{C-type} \\
        1 - \Sigma                & \text{if}\; i\;\in\; \text{O-type} \\
        0.5 + \mu (0.5 - \lambda) & \text{if}\; i\;\in\; \text{SE-type} \\
        0.5 + \mu (\lambda - 0.5) & \text{if}\; i\;\in\; \text{ST-type}
    \end{array} \right.
\end{equation}

Since satisfaction carries over from one day to another, we can state that $S_{i,t} = S_{i,t-1}$ at the beginning of a new time step in the simulation.
Should $S_{i,t-1}$ deviate positively (negatively) from $S_{i}^{0}$, it is reduced (increased) by 1\% of its value, as formulated in Equation \ref{eq: satisfaction-offset}.\footnote{
    It is also conceivable to model satisfaction recovery in a non-linear fashion such that only the \textit{offset} from base satisfaction would be reduced by 1\%.
    A possible formalisation would be where greater deviations from base satisfaction are reduced faster and a total recovery back to $S_{i}^{0}$ is made impossible, such that $S_{i,t} = S_{i,t-1} - \frac{S_{i,t-1} - S_{i}^{0}}{100}$.
    Regardless of the chosen implementation, the restriction of $S_{i,t} \in [0,1]$ shall always hold.
}

\begin{equation}\label{eq: satisfaction-offset}
    S_{i,t} =
    \left\{ \begin{array}{ll}
        0.99 \; S_{i,t-1} & \text{if}\; S_{i,t-1} > S_{i}^{0} \\
        1.01 \; S_{i,t-1} & \text{if}\; S_{i,t-1} < S_{i}^{0}
    \end{array} \right.
\end{equation}

During each period, the management observes a random subset of workers and controls for excessive shirking levels.
The management checks a subset of randomly drawn employees $ETC \subset N$ with cardinality $|ETC| = \Sigma \cdot n$.
We assume that the management willingly accepts a certain amount of shirking activity ($s^{max}$) because it is inevitable to some degree and might even be beneficial \citep{Vermeulen.2000, Campbell.2019}.
This threshold might be subject to various considerations such as the firm's desired revenue or profit margin, management values, or just the observed behaviour of employees.
For the time being, the simulations conducted with this second model extension will use a constant value of one-tenth of the available working time of all agents, i.e. $s^{max} = \tau / 10$.
Thus, when checking on employees, management deems their shirking levels to be reasonable as long as $s_{i,t} \leq s^{max}$.

Since receiving a warning from superiors is generally a negative experience, employees become more dissatisfied after getting caught shirking too much, thus lowering their productivity.
To which extent getting caught impacts agents' degree of satisfaction $S_{i,t}$ might depend on agents' value types \citep{Chatman.1991}, however, we model the impact on satisfaction after receiving a verbal warning in the same manner for all agent types.
Thus, a verbal warning will reduce employee satisfaction by an arbitrary \textit{shock of being caught} ($\eta = [0,1]$) such that $S_{i,t} = S_{i,t}(1 - \eta)$.
The simulation results discussed in this paper have used a constant $\eta = 0.05$.

If workers get caught \textit{for the third time} shirking more than accepted, the management will issue a written warning ($ww_{i,t}$) signalling that repeating such behaviour might result in some form of punishment.\footnote{
    That being said, there is no form of consequence or punishment implemented in the presented model.
    Therefore, this provides an intriguing venue for future research, as for example in a model dealing with hiring-firing mechanisms and their impact on the labour market.
}
The warnings have two effects:
    (i) the worker might shirk less in the future for fear of bad consequences; hence \textit{individual} deviations from the shirking norm are modelled along with the type-specific ones; 
    (ii) workers get \textit{dissatisfied} with their work.
This implies the existence of an optimum degree of monitoring for which the positive deviation from the shirking norm is minimised while keeping a high employee satisfaction (and thus productivity).

\begin{figure}[t]
    \centering
    \includegraphics[width=0.7\textwidth]{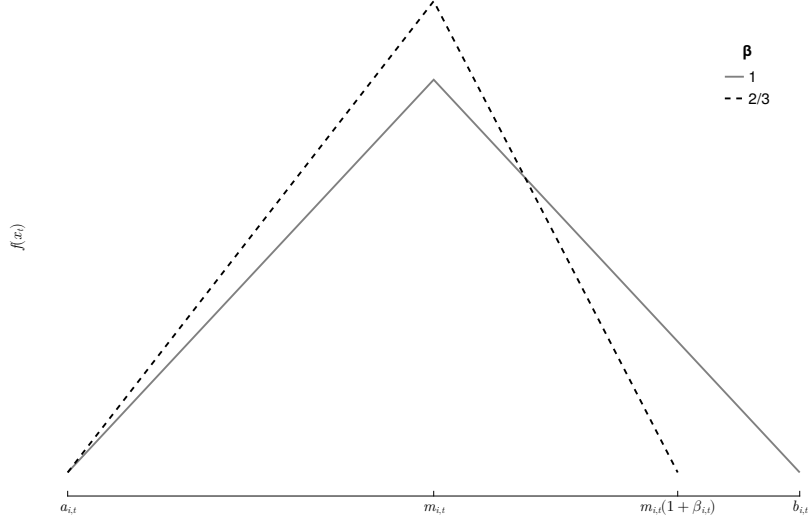}
    \caption{Density function of a triangular distribution for shirking behaviour with $\beta_{i,t} \; = \; \sfrac{2}{3}$. Adaptation of \cite{Roos.2022}'s triangular distributions of agents' stochastic behaviour. The equations defining the parameters of the triangular distribution can be found in Table \ref{tab: list_eq}. Source: Authors' own illustration.}
    \label{fig: beta}
\end{figure}

Letters of reprimand are written in reaction to management's actual observations of shirking behaviour, hence, receiving one reduces workers' future positive deviations from the shirking norm.
Formally we model this with an individual-specific scaling factor $\beta_{i,t}$, with $\beta_{i,t} = 1$ as its default state, which alters the upper bound of the triangular distribution used for individual decision making\footnote{
    The original upper bound was $b_{i,t} = s_{i,t}^{*} (1 + \delta_{i})$, as can be seen on row number 10 of Table \ref{tab: list_eq} in Appendix \ref{sec: app-1}.
}, see Equation \ref{eq: shirking-upper-bound}.
Figure \ref{fig: beta} provides an example of what would happen for a $\beta_{i,t}$ of $\sfrac{2}{3}$ (red line) versus the baseline case (black line) assumed in \cite{Roos.2022}.

\begin{equation}\label{eq: shirking-upper-bound}
    b_{i,t} = s^{*}_{i,t} (1 + \beta_{i,t\phantom{i}}\delta_{i})
\end{equation}

Changes to $\beta_{i,t}$ are assumed to have a persistent effect which gradually decreases over time.
Thus, the more time has passed since the last written warning was received, the less an employee's value-based behaviour is modified.
To capture this, written warnings are modelled as a finite set that takes record of the steps at which the agent has been caught shirking more than acceptable for the third time: $WW_{i,t} = \{ww_{1}, ww_{2}, \dots, ww_{n} \}$.
If the set is non-empty, employees will permanently alter their shirking behaviour according to \textit{how long ago} the last warning ($ww_{n} : ww_{n}\; \in\; WW_{i,t} \neq \emptyset$) was received and \textit{how many} warnings ($|WW_{i,t}|$) were recorded overall, see Equation \ref{eq: beta}.

\begin{equation}\label{eq: beta}
    \beta_{i,t}= 
    \left\{ \begin{array}{lll}
    1 - \frac{|WW_{i,t}|}{3} + \frac{|WW_{i,t}|}{3} \; \frac{t - x_{n}}{t} & \text{if}\; 0 \leq |WW_{i,t}| < 3\\
    0.0 + 1.0 \; \frac{t - x_{n}}{t} & \text{otherwise}\\
    \end{array} \right.
\end{equation}\vspace{-0.5em}
    
Instead of just being verbally admonished to shirk less, receiving a written warning is an important formal signal of management control which results in a bigger impact on employee satisfaction.
We chose a factor of three for the simulations in the work at hand.
Hence, after receiving a written warning, the affected agent's satisfaction is reduced to a fraction of its former value, such that $S_{i,t} = S_{i,t} (1 - 3 \eta)$.

\subsection{Endogenous management strategies}\label{subsec: end_management}

Differently from the static and exogenous management assumptions in \cite{Roos.2022}, both monitoring and incentives are dynamic and endogenous here.
The management tracks key company benchmarks, namely average company output ($\bar O_{t}$) as well as the average shirking ($\bar s_{t}$) and cooperative times ($\bar c_{t}$) of observed workers over the past $x$ periods.
The average observed shirking and cooperative times are defined respectively as

\begin{equation}\label{eq: mean-obs-shirk}
    \bar s_{t} = \frac{1}{n} \sum_{i}^{ETC_{t}} s_{i,t}
\end{equation}
and
\begin{equation}\label{eq: mean-obs-coop}
    \bar c_{t} = \frac{1}{n} \sum_{i}^{ETC_{t}} c_{i,t} 
\end{equation}

where $ETC_{t}$ is the now endogenous subset of observed workers with cardinality $|ETC_{t}| = \Sigma_{t-1} \cdot n$.\footnote{
    Please note that in the previous extension (Section \ref{subsec: adaptive-behaviour}) the cardinality of the subset of observed workers was exogenous, as $\Sigma$ was not responding to any company benchmarks.
}
Management judges recent developments based on the preset goals of expected group output (Equation \ref{eq: EGO}), the exogenous degree of task interdependence $\kappa$ (Equation \ref{eq: output_eq}), and maximum acceptable shirking time (Equation \ref{eq: max-shirking}).

In contrast to the fixed value chosen in the previous extension, the maximum acceptable shirking time is now endogenised as $s_{t}^{max}$ which reflects the management's adaptive expectations regarding the usual and necessary work efforts of the firm employees.
Considering the deliberate absence of any external (e.g. market-related) factors in our model that might influence management behaviour, we propose that the maximum accepted shirking level is modelled in a similar fashion to how it has been done for other social norms.
Thus, $s_{t}^{max}$ can be understood as the shirking norm perceived by the management and depends on both its previous value ($s_{t-1}^{max}$) and the mean shirking behaviour of the observed agents on the previous day ($\bar s_{t-1}$).
How quickly the management adapts $s_{t}^{max}$ again depends on the exogenous parameter $h$ previously used in Equations \ref{eq: shirking-norm} and \ref{eq: cooperation-norm}.\footnote{
    In a model including value-based management, these changes in maximum acceptable shirking time could be further endogenised with theory-driven behavioural rules.
}
Instead of peer influence as in the case of agents, changes to $s_{t}^{max}$ depend on those agents that the management has controlled on the previous day.
Note that the reference point is shifted from an individual to a top-down aggregate view (see the second column of Table \ref{tab: list_eq} in Appendix \ref{sec: app-1}).

\begin{equation}\label{eq: max-shirking}
    s_{t}^{max} = (1 - h) \; s_{t-1}^{max} + h \; \bar s_{t-1}
\end{equation}

The management can now infer an expected group output $EGO_{t}$, which is the maximum of the Cobb-Douglas type production function under the constraint $s_{t} = s_{t}^{max}$.
Let us denote employees' available time out of the maximum acceptable shirking threshold with $\alpha_{t} = \tau - s_{t}^{max}$, $EGO_{t}$ is then defined as follows.

\begin{equation}\label{eq: EGO}
    EGO_{t} = \left[\alpha_{t}(1-\kappa)\right]^{(1-\kappa)} \cdot (\alpha_{t}\kappa)^{\kappa}
\end{equation}

Monitoring and incentive strategies are updated in a pre-determined interval according to a \textit{strategy update frequency} parameter ($suf \in \mathbb{N}$).
The future degree of corporate monitoring ($\Sigma_{t} \in [0,1]$) is determined as in Equation \ref{eq: sigma_endog} where $sui \in \mathbb{R}^{+}$ is an exogenous \textit{strategy update intensity} parameter.
We have chosen $suf = \{1,30,180,365\}$, $sui = \{\sfrac{1}{60}, \sfrac{1}{20}, \sfrac{3}{10}, \sfrac{73}{120}\}$, and $x = suf$ for the results discussed in Section \ref{sec: discussion}.

\begin{equation}\label{eq: sigma_endog}
    \Sigma_{t} = 
    \left\{ \begin{array}{ll}
    (1 + sui) \Sigma_{t-1} & \text{if}\; \frac{1}{x}\sum\limits_{t-x}^{t-1} \bar s_{t} > s_{t}^{max}\\
    (1 - sui) \Sigma_{t-1} & \text{if}\; \frac{1}{x}\sum\limits_{t-x}^{t-1} \bar s_{t} \leq s_{t}^{max}\\
    (1 - \frac{\bar O_{t}}{EGO_{t}}) sui  & \text{if}\; \Sigma_{t-1} = 0 \; \wedge \; \frac{1}{x}\sum\limits_{t-x}^{t-1} \bar O_{t} < EGO_{t}
    \end{array}
    \right.
\end{equation}

Therefore, the management becomes more (less) controlling when the average observed shirking ($\bar{s}_{t}$ ) of the current set of monitored employees ($\forall\; i \in ETC_{t}$) exceeds (stays within) the maximum predefined threshold $s_{t}^{max}$.
As can be noted from Equation \ref{eq: sigma_endog}, we also account for a special case which takes place when the management has adopted a fully trusting strategy in the previous period, i.e. when $\Sigma_{t-1} = 0$.
When this event occurs, it is reasonable to conceive the management as indifferent to employees' shirking attitudes.
In this case, the firm would instead anchor any monitoring decisions to the average company output ($\bar O_{t}$) such that $\Sigma_{t}$ would increase proportionally to how far away $\bar O_{t}$ was from expected group output ($EGO_{t}$).

While monetary incentives are assumed to have positive steering effects on employee motivation \citep{Gerhart.2017}, the management should keep the amount of financial rewards as low as feasible as it contributes to the overall costs of the firm.
Wages are assumed to be sticky to some extent, as reflected in Equation \ref{eq: mu_endog}.
The management increases (decreases) the amount of financial rewards when the average company output ($O_{t}$) is below (above) the management's expected group output.
If the company benchmark – namely $EGO_{t}$ – is reached, we assume that the management has no incentive to alter the amount of rewards.

\begin{equation}\label{eq: mu_endog}
    \mu_{t} = 
    \left\{ \begin{array}{ll}
    (1 + sui) \mu_{t-1} & \text{if}\; \frac{1}{x}\sum\limits_{t-x}^{t-1} \bar O_{t} < EGO_{t} \\
    (1 - sui) \mu_{t-1} & \text{if}\; \frac{1}{x}\sum\limits_{t-x}^{t-1} \bar O_{t} > EGO_{t} \\
    \mu_{t-1} & \text{otherwise} \\
    \end{array}\right.
\end{equation}

When the management observes a subset of employees ($ETC_{t}$), it also gathers information about the amount of time they have devoted to cooperative activities ($\bar c_{t}$).
The management shifts to a higher (lower) degree of competitive rewards when the desired amount of time allocation to cooperation ($\kappa \cdot \alpha_{t}$) is (not) achieved.
By doing so, we also account for mixed PFP schemes ($\lambda_{t} \in [0,1]$), i.e. schemes that combine collective and individual rewards, which represent the most common type of rewards used in real-world scenarios \citep{Nyberg.2018}.

\begin{equation}\label{eq: lambda_endog}
    \lambda_{t} = 
    \left \{ \begin{array}{ll}
    (1 + sui) \lambda_{t-1} &\text{if}\; \frac{1}{x}\sum\limits_{t-x}^{t-1} \bar c_{t} < \kappa \cdot \alpha_{t} \\
    (1 - sui) \lambda_{t-1} &\text{if}\; \frac{1}{x}\sum\limits_{t-x}^{t-1} \bar c_{t} > \kappa \cdot \alpha_{t} \\
    \lambda_{t-1}  & \text{otherwise} \\
    \end{array} \right.
\end{equation}

Equations \ref{eq: sigma_endog}, \ref{eq: mu_endog}, and \ref{eq: lambda_endog} could result in values below or above the parameter boundaries of $[0,1]$.
In these cases, $\Sigma_{t}$, $\mu_{t}$ or $\lambda_{t}$ will be rounded to the nearest possible value inside the interval.
Further, any changes to the management style ($\Sigma_{t}$, $\mu_{t}$ or $\lambda_{t}$) also induce an update of the base satisfaction of agents $S_ {i}^{0}$ in the same way as described in Equation \ref{eq: base-satisfaction}.

\section{Simulations and results}\label{sec: results}

Our main research question is about the potential effect of corporate culture on the profit differentials of otherwise similar firms.
To shed light on this, the current section presents our findings from the conducted agent-based simulations by focusing on three aspects identified to impact profitability: (i) the frequency of changes in management decisions, (ii) the influence of employees' homophily in interactions, and (iii) the role of job satisfaction.

A detailed description of the agent-based simulation algorithm can be found in Figure \ref{fig: flowchart} in Appendix \ref{sec: app-2}.
The simulations have been run for $3650$ time steps (i.e. $10$ years) with a stable workforce.
The presented results are mean aggregates over $100$ uniquely seeded replicate runs for each parameter constellation (= scenario).
All initial model parameters are summarised in Table \ref{tab: parameters} in Appendix \ref{sec: app-1}.\footnote{
    The Julia code to recreate all of the simulation results and visualisations is available online (see \href{https://git.noc.ruhr-uni-bochum.de/vepabm/firm-as-cas}{here}) which comes with a thorough explanation of how to run it.
}
The nine scenarios previously used by \cite{Roos.2022} have become meaningless here with the endogenisation of the management strategy.
As such, all further simulations will start from the same neutral management strategy which equals to the previously used Base scenario.\footnote{
    The chosen starting values for model parameters can potentially influence the simulation results through path-dependency and lock-in effects.
    Appendix \ref{sec: app-3} provides the results of our conducted sensitivity analyses regarding the initial values of strategy parameters $\Sigma$, $\mu$, and $\lambda$ as well as four different initialisation methods for employee time allocation.
}
As outlined in the model description (see Section \ref{sec: model}), the management strategy regarding monitoring efforts and implemented pay for performance scheme now changes over time and depends on the chosen strategy update intensity ($sui$) and frequency ($suf$).
Hence, we introduce four new scenarios in Table \ref{tab: EM-scenarios} that modulate these parameters with which the management deterministically reacts to changes in the observed firm variables.\footnote{
    While both $suf$ and $sui$ are modulated, the ratio of $\sfrac{sui}{suf}$ is held constant to keep the amount of scenarios low.
    This assumption can be relaxed for more in-depth analysis of these model parameters.
}

\begin{table}[H]
    \centering
    \begin{tabular}{lcc}
        \toprule
        Name & $suf$ & $sui$ \\
        \hline \hline
        Daily & 1 & 1/600 \\
        Monthly & 30 & 1/20 \\
        Biannually & 180 & 3/10 \\
        Yearly & 365 & 73/120 \\
        \bottomrule
    \end{tabular}
    \caption{Varying strategy update intensity ($sui$) and frequency ($suf$) combinations in four scenarios for use with the endogenous management extension.}
    \label{tab: EM-scenarios}
\end{table}

\begin{figure}[t]
    \centering
    \includegraphics[width=\textwidth]{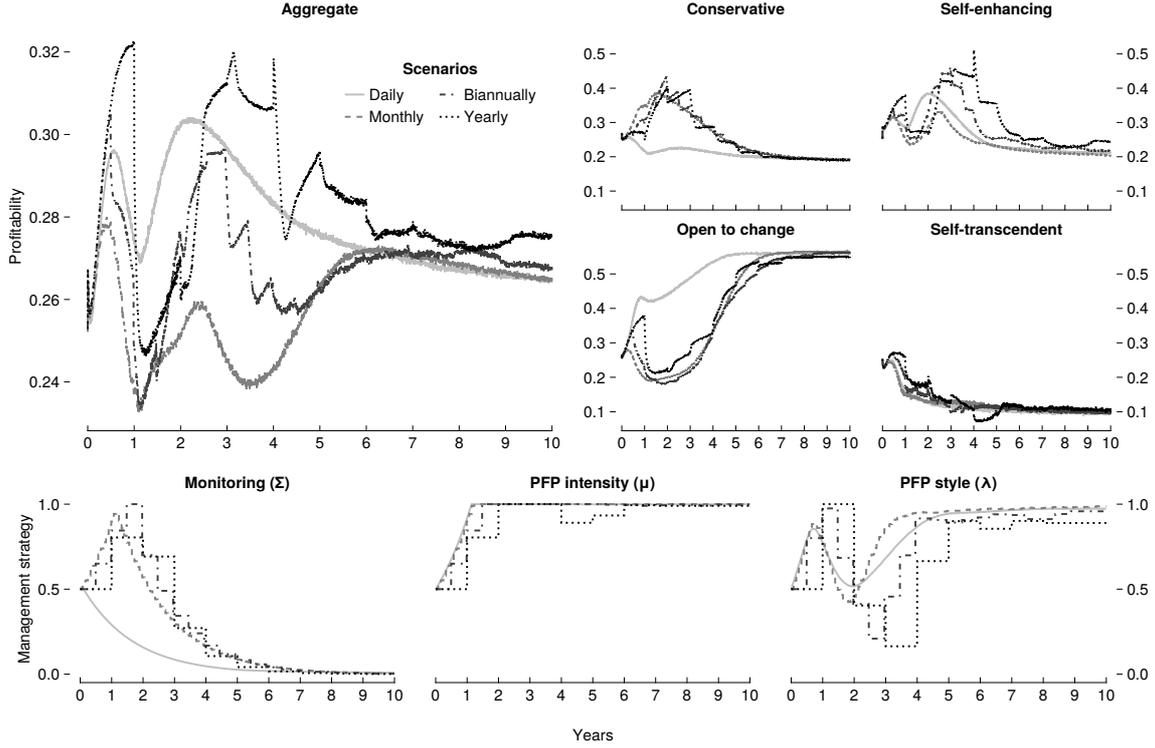}
    \caption{Firm profitability (top row) and management strategy (bottom row) over time across four scenarios. Top left plot shows aggregate profitability across the whole work force. Top right plots show profitability broken down into the four higher-order value groups. Three plots in bottom row show the development of the management strategy parameters. Source: Authors' own illustration.}
    \label{fig: profitability}
\end{figure}

Figure \ref{fig: profitability} displays the firms profitability over time across four scenarios with varying degrees of strategy update frequency and intensity (see Table \ref{tab: EM-scenarios} for details).
Profitability has been formalised as the ratio of sum of output to sum of rewards.\footnote{
    Note that this ratio is of deeply theoretical nature as the model at hand does not have a market to convert output into money.
    As such, the relative differences in profitability allow for a discussion of corporate culture as a potential source of profit differentials between otherwise equal firms.
    Therefore, this abstraction to relate units of output to unspecified monetary units of reward payments seems sufficient for the aims of this paper.
}
The main plot in the top left subfigure shows this from an aggregate perspective, taking into account the output and rewards of all firm employees.
The top right contains four subfigures providing a more fine-grained view on the profitability of each of the four value groups.
The evolution of the management strategy according to the three parameters monitoring ($\Sigma$), intensity ($\mu$), and type ($\lambda$) of implemented PFP scheme can be tracked in the bottom row of Figure \ref{fig: profitability}.

The top left plot shows that changes in profitability get more erratic with decreasing strategy update frequency whereas more incremental updates lead to a smoother transition over time.
This allows the firm's management to react faster to new business insights and closely adapt their currently employed strategy in accordance with the underlying heuristics.
After one year the Yearly scenario brings the best results in terms of profitability, which points towards the positive impact of a stable environment that allows social norms to manifest and spread among the employees.
However, this changes rapidly in the following year where drastic modifications of the management style (increased monitoring due to higher than acceptable observed shirking) lead to severe drops in profitability.
Although Conservative agents react positively to this change, even leading to a short-lived rise in profitability for these employees, the decline in aggregate profitability can be observed in Figure \ref{fig: profitability} across all value-types from the second year onward, where only Open-to-change agents' profitability shows a U-shaped recovery.

One noteworthy outlier is the profitability of Conservative and Open-to-change agents in the Daily scenario, implying that small and frequent updates lead to (un)favourable outcomes for employees in these higher-order value group.
Here, the management can constantly observe the shirking of a subset of employees and adapt its expectations (i.e. $s_{t}^{max}$ and therefore also $EGO_{t}$) to what has actually happened over the past day.
The resulting early drops in monitoring efforts $\Sigma$ significantly increase the satisfaction levels (i.e. productivity) of Open-to-change agents.\footnote{
    The inverse effect of more monitoring can also be observed in the evolution over time of Monthly, Biannually, and Yearly scenarios.
}
Even though the model has been built under the assumption of behavioural symmetry between employees of opposing higher order values, Conservative agents do not completely mirror the reactions of Open-to-change agents because their increasingly productive behaviour also manifests itself in the social norms perceived by others.
Furthermore, any reduction in monitoring also lowers the amount of observed employees each day, thus leading to less verbal and written warnings, and ultimately resulting in higher satisfaction across the whole population.
In the Daily scenario, this effect counters the negative impact on satisfaction of Conservative employees throughout years one to three and even leads to small increases in their profitability.
Yet the long-run trend towards very low degrees of monitoring throughout all four scenarios eventually overshadows these gains and lead to convergence of Conservative and Open-to-change employees' profitability around $0.19$ and $0.56$ respectively.

Although also positively affected by the reduction in warnings issued by the management, the evolution of Self-enhancing and Self-transcendent agents is driven by different influential factors.
With respect to the intensity of implemented incentive schemes, the four scenarios paint a very similar, albeit slightly time-lagged, picture.
After two years at most, they all lead to maximum $\mu$, leaving this parameter around this level until the end with only a short-lived dip in PFP intensity in years four to six of the Yearly scenario.
As laid out in Equation \ref{eq: mu_endog}, changes to the amount of incentives paid depend on management's expectations regarding expected group output which is starkly influenced by what is observed over time by the management as normal shirking behaviour.
Indeed, the results hint at those expectations being practically unachievable under most simulation settings.
The only exception is a period of two years in the Yearly scenario where the sustained productivity of Self-enhancing employees and the rapidly increasing productivity of Open-to-change employees contribute to levels of output that are high enough to warrant a reduction in monetary incentives.
Still, it is important to note that the amount of paid incentives has a strong impact on the calculation of profitability, ultimately adding to the explanation of the ongoing side-/downward trend in aggregate profitability after year six (and earlier when looking at the separate higher order value groups).

Changes to the type of implemented PFP scheme occur in the first half of the simulation period and cause positive/negative reactions from Self-enhancing/Self-transcendent agents.
However, the positive impact on the former group's behaviour is diminished by their peers from other value groups, thus countering the development of potentially more profitable social norms.
This kind of mitigation cannot be observed for Self-transcendent employees which is caused by their social networks exhibiting high degrees of homophily\footnote{
    Homophily is defined as the weighted share of agent $i$'s peers who belong to the same value group as $i$.
    The weights are determined by the current connection strength at time $t$ between agents $i$ and $j \; \forall \; j \in P_{i,t}$.
} at approximately twice the levels of all other value groups in the long run.
As such, this management decision is overshadowed by the declining influence of Self-transcendent employees on the social norms across the whole firm.
Subsequently the decision to lower $\lambda$ is reverted from years 2 (Daily) to 4 (Yearly) onward, eventually remaining at high levels above $0.9$ again.

\begin{figure}[t]
    \centering
    \includegraphics[width=\textwidth]{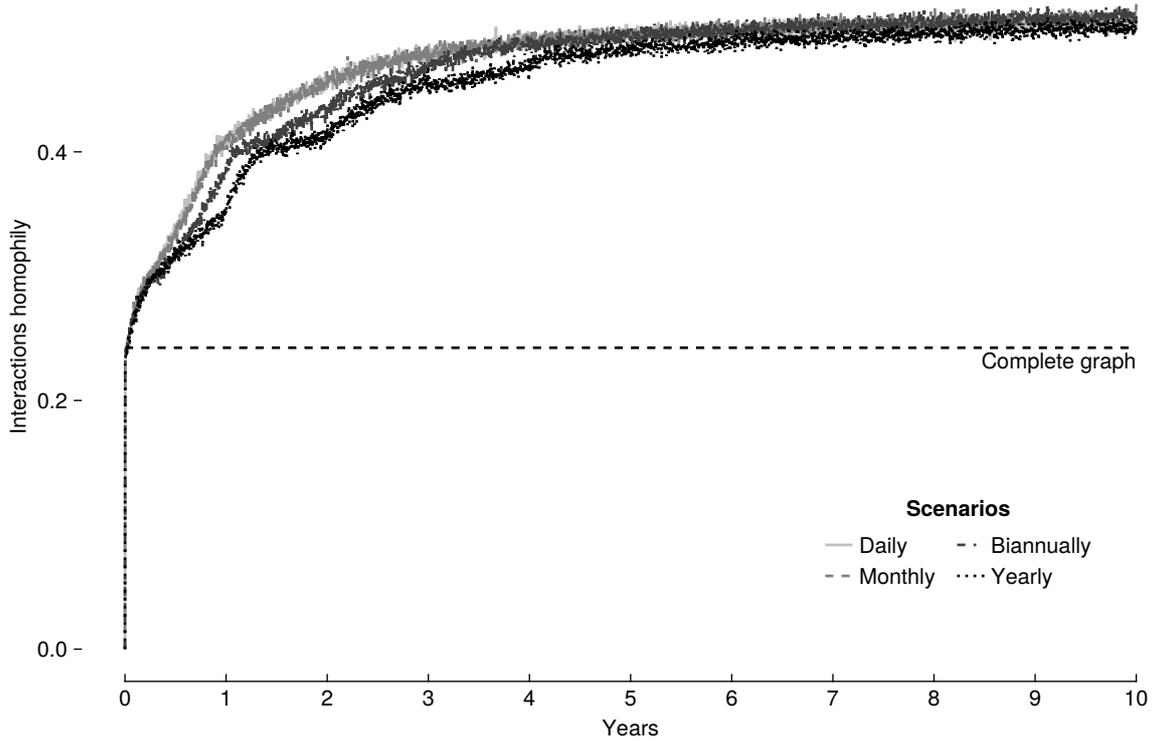}
    \caption{Interactions homophily in the endogenous social network across four scenarios. Left plot shows aggregate interactions homophily across the whole work force. Right plots show interactions homophily broken down into the four higher-order value groups. Source: Authors' own illustration.}
    \label{fig: homophily-interactions}
\end{figure}

Figure \ref{fig: homophily-interactions} provides insights on how homophily in interactions between the employees changes over time.
The main plot on the left side depicts the average homophily of all agents' interactions in the endogenous social network across the four scenarios and also provides a dashed horizontal line as a reference case with an unweighted complete graph.
The four subplots on the right side are again divided by the higher-order value types and display their mean interactions homophily for inter-group comparisons.
While it is evident that differences exist between all four employee types, Self-transcendent agents reach severely higher degrees of homophily ($0.80 - 0.85$) than the three other groups.
Since the probability for two agents to interact depends on their activity similarity, the stronger deviations from social norms allow Open-to-change agents to consistently achieve the most inter-group interactions at the end of the four simulated scenarios ($0.37 - 0.38$).
Conservative ($0.39 - 00.41$) and Self-enhancing ($0.41 - 0.46$) employees find themselves in the middle.
The explanation for the wider spread of the latter group can be found in the different intensity of competitive incentives across the four scenarios (cf. bottom right plot in Figure \ref{fig: profitability}) which leads to temporary boosts to activity similarity in this value group.
These findings suggest that even short-lived changes in management strategy can have a lasting effect on the firm's network and by that consequently also affect profitability.

\begin{figure}[t]
    \centering
    \includegraphics[width=\textwidth]{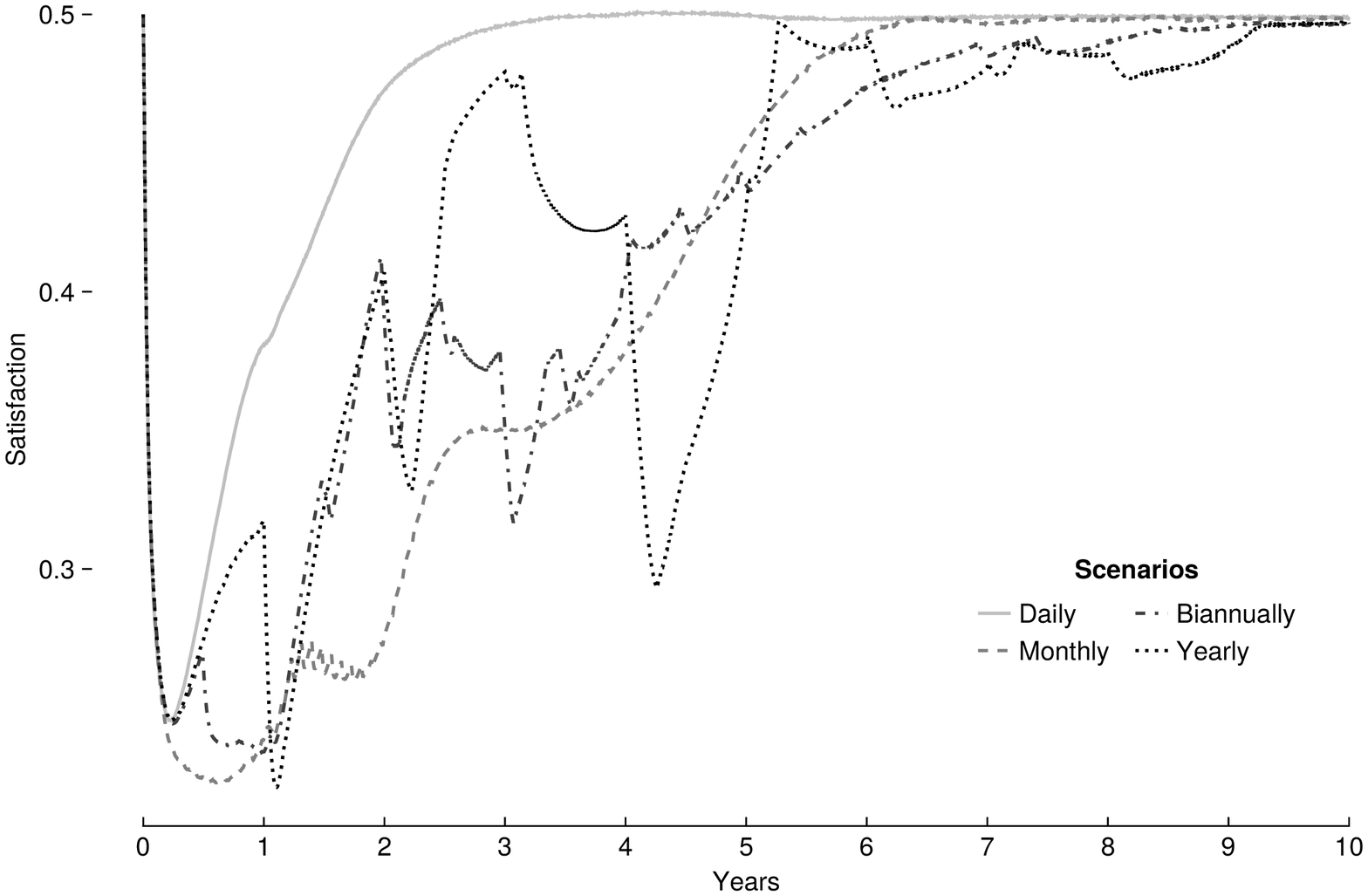}
    \caption{Job satisfaction across four scenarios. Left plot shows aggregate satisfaction across the whole work force. Right plots show satisfaction broken down into the four higher-order value groups. Source: Authors' own illustration.}
    \label{fig: satisfaction}
\end{figure}

Comparing the satisfaction (see Figure \ref{fig: satisfaction}) and profitability curves of the four agent groups reveals high similarities of their evolution for almost all employees except those belonging to the Self-transcendent group.
The correlation values reported in Table \ref{tab: correlations} show that bivariate correlations between satisfaction and profitability are strong for Conservative, Open-to-change, and Self-enhancing agents ($SP \geq 0.879$), which suggests that job satisfaction is indeed a positive influential factor for these employees.
The low correlation of Self-transcendent agents is an indicator that there are cases in which job satisfaction is indeed high with no opportunity to translate this into high levels of output and/or profitability.
This is due to their generally unproductive allocation of time with too much emphasis on cooperative activities and the accompanying social separation from the rest of the firm's employees which in combination lead to low profitability of Self-transcendent agents.
However, for all higher order value groups a lower strategy update frequency generally implies a more positive correlation between satisfaction and profitability.

\begin{table}[H]
    \centering
    \begin{tabular}{llccc}
        \toprule
        Value group & Scenario & $SP$ & $HP$ & $SH$ \\
        \hline \hline
        \multirow{4}*{Conservative}
            &   Daily       &  \phantom{-}0.879 &  -0.703           &  -0.817 \\
            &   Monthly     &  \phantom{-}0.977 &  -0.236           &  -0.291 \\
            &   Biannually  &  \phantom{-}0.978 &  -0.256           &  -0.238 \\
            &   Yearly      &  \phantom{-}0.970 &  -0.274           &  -0.225 \\
        \cmidrule{2-5}
        \multirow{4}*{Open-to-change}
            &   Daily       &  \phantom{-}0.943 &  \phantom{-}0.948 &  \phantom{-}0.899 \\
            &   Monthly     &  \phantom{-}0.998 &  \phantom{-}0.831 &  \phantom{-}0.818 \\
            &   Biannually  &  \phantom{-}0.998 &  \phantom{-}0.843 &  \phantom{-}0.842 \\
            &   Yearly      &  \phantom{-}0.994 &  \phantom{-}0.860 &  \phantom{-}0.879 \\
        \cmidrule{2-5}
        \multirow{4}*{Self-enhancing}
            &   Daily       &  \phantom{-}0.982 &  -0.126           &  -0.156 \\
            &   Monthly     &  \phantom{-}0.927 &  -0.350           &  -0.494 \\
            &   Biannually  &  \phantom{-}0.973 &  \phantom{-}0.013 &  \phantom{-}0.053 \\
            &   Yearly      &  \phantom{-}0.954 &  \phantom{-}0.071 &  \phantom{-}0.158 \\
        \cmidrule{2-5}
        \multirow{4}*{Self-transcendent}
            &   Daily       &  -0.855           &  -0.977           &  \phantom{-}0.883 \\
            &   Monthly     &  -0.838           &  -0.966           &  \phantom{-}0.902 \\
            &   Biannually  &  -0.651           &  -0.962           &  \phantom{-}0.761 \\
            &   Yearly      &  -0.268           &  -0.931           &  \phantom{-}0.519 \\
        \cmidrule{2-5}
        \multirow{4}*{Average}
            &   Daily       &  -0.141           &  -0.220           &  \phantom{-}0.937 \\
            &   Monthly     &  \phantom{-}0.602 &  \phantom{-}0.138 &  \phantom{-}0.760 \\
            &   Biannually  &  \phantom{-}0.006 &  -0.090           &  \phantom{-}0.843 \\
            &   Yearly      &  \phantom{-}0.023 &  -0.110           &  \phantom{-}0.793 \\
        \bottomrule
    \end{tabular}
    \caption{Bivariate Pearson correlations between satisfaction and profitability ($SP$), interactions homophily and profitability ($HP$), and satisfaction and interactions homophily ($SH$). Results have been truncated to three digits.}
    \label{tab: correlations}
\end{table}

Homophily has weak explanatory value for the profitability of Self-enhancing agents and relatively low, although firmly negative, correlations for Conservative agents.
One exception is the Daily scenario where the early reduction of monitoring leads to a vastly divergent result and by that pushes the correlation coefficient of satisfaction and profitability for Conservative agents further into negative territory.
The interaction homophily of Open-to-change (Self-transcendent) agents follows their profitability in more pronounced ways, exhibiting strong positive (negative) correlations that decrease in intensity with lower strategy update frequency.
Satisfaction and interaction homophily evolve in similar ways for both Open-to-change and Self-transcendent agents and show only low signs of correlation for Self-enhancing agents.
However, there is a negative correlation for Conservative agents implying that higher satisfaction levels are accompanied by lower homophily in their interactions.
These observations suggest that a management might want to implement measures that increase the embeddedness of Self-transcendent agents in the broader firm population (thus lowering their homophily levels) while at the same time such measures are likely to be relatively ineffective for Self-enhancing agents.
For Conservative, Open-to-change, and Self-enhancing employees the effects of high job satisfaction on profitability are likely stronger than the impact of their personal social network, suggesting that a sensible management would instead cater to this by implementing measures that raise their general job satisfaction.

The last four rows of Table \ref{tab: correlations} show how the variables correlate with each other when using average values computed from all firm employees.
In the Biannually and Yearly scenarios, both satisfaction and interactions homophily show only very slight correlations to firm profitability which suggests that these aspects of corporate culture play a relatively minor role in scenarios with slow-moving management strategies.
This effect is more pronounced in the Monthly scenario ($SP = 0.602$) which is also the only case where interactions homophily has a positive correlation coefficient with profitability ($HP = 0.138$).
Quite contrarily, daily management strategy updates lead to a situation in which satisfaction and interactions homophily are strongly positively correlated ($SH = 0.937$) and higher values are accompanied by lower profitability.\footnote{
    While these results are somewhat unexpected and might even warrant a more in-depth examination and discussion, it has to be duly noted that the Daily scenario is an extreme edge case.
    It implies perfect willingness and ability of both employees and management to adapt to an ever-changing work environment and does not incorporate any cost of changing strategies (which arguably becomes more important with higher strategy update frequencies).
    Issues related to this aspect have for instance been discussed by \cite{Cremer.1993}.
}
To summarise, three out of four scenarios show a more or less moderate positive correlation between satisfaction and profitability which qualitatively fits the findings in the empirical literature \citep{Judge.2001, Fisher.2003}.

Going back to Figure \ref{fig: profitability}, after some time the four scenarios have reached states of slowly declining profitability past $6$ simulated years which persist until the end of the simulation.
Yearly strategy updates yield the highest profitability at approximately $0.2751$ after ten years.
The other three scenarios reached profitability levels closer to each other with Biannually at $\approx 0.2675$, Monthly at $\approx 0.2653$, and Daily at $\approx 0.2647$.
Thus, it becomes apparent that (i) fluctuations in profitability increase with less frequent changes in management style, (ii) achieved levels of profitability are higher under a less adaptive management, and (iii) management expectations regarding output and normal degrees of shirking play a crucial role in the long-term profitability of firms.

\begin{figure}[t]
    \centering
    \includegraphics[width=\textwidth]{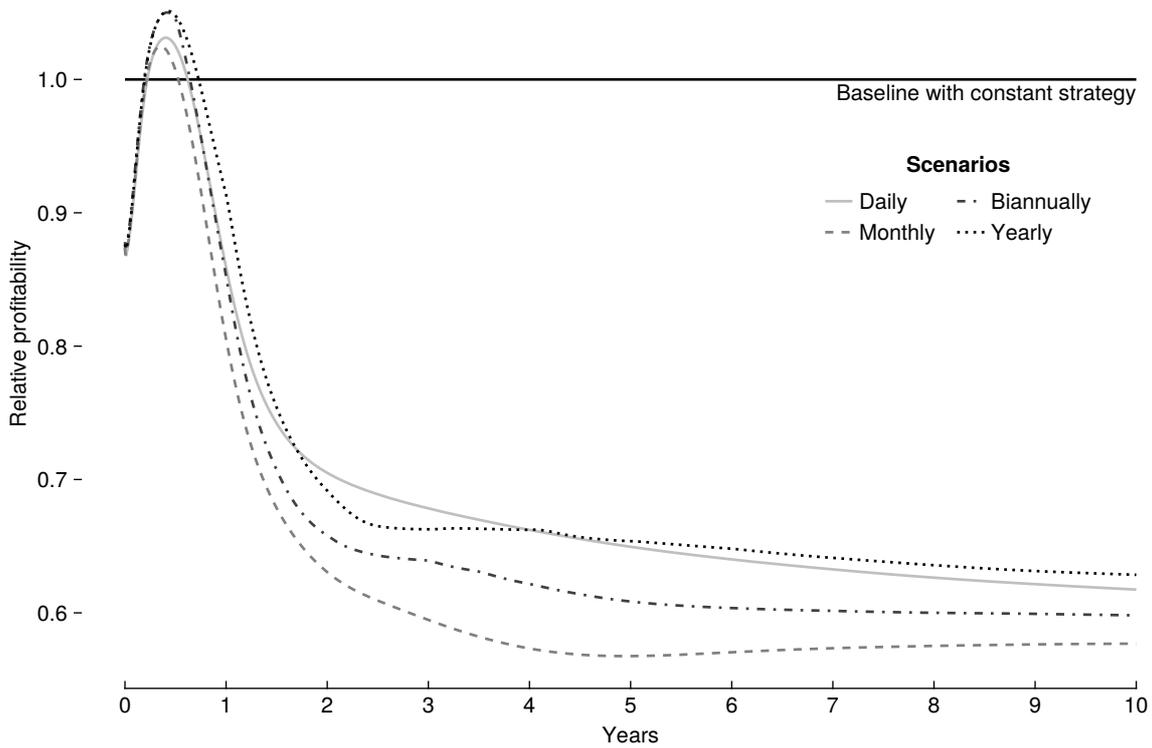}
    \caption{Cumulated firm profitability over time for each of the four scenarios divided by the cumulated firm profitability of a baseline scenario with a neutral and constant management style (black dashed horizontal line). Source: Authors' own illustration.}
    \label{fig: relative-profitability}
\end{figure}

Although the differences between the scenarios are clearly visible, their endpoints are relatively close to each other while also only providing a snapshot of the current profitability at any given time.
Cumulated profitability over time paints another picture, however, as it takes into account the accumulated profits in each scenario's pathway which can then be compared to a neutral baseline scenario\footnote{
    The model parameters shaping the management style of this scenario are $\Sigma = 0.5$, $\mu = 0.0$, and $\lambda = 1.0$ which is identical to the Base scenario used in \cite{Roos.2022}.
} without any changes in management style.
We can see in Figure \ref{fig: relative-profitability} that the four lines at first follow similar trajectories but indeed deviate from each other's paths over the long run.

\begin{table}[H]
    \centering
    \begin{tabular}{r|ccccc}
        Scenario & Base & Yearly & Daily & Biannually & Monthly \\
        Relative profitability & 100.00 & 62.85 & 61.75 & 59.81 & 57.69 \\
    \end{tabular}
    \caption{Relative profitability of the four scenarios at the end of the simulation in percent, sorted in descending order from left to right. This is measured as the relation of their cumulated firm profitability values to that of the baseline scenario without endogenised management decision making.}
    \label{tab: relative-profitability}
\end{table}

Considering only the end of the simulations, Table \ref{tab: relative-profitability} provides some insight on the fact that the differences in cumulated profits are deeply ingrained in the emerging corporate culture that is shaped by social norms and the frequency of strategic management decisions.
The Yearly scenario yields the highest cumulated profitability in the long run even though slow changing strategies may lead to managerial overreaction due to lower ability to adapt to new insights in the short term.
Daily and biannual changes in management strategies lead to medium levels of cumulated profitability whereas pursuing monthly strategy changes consistently leads to the worst performance among the four scenarios.

However, it has to be noted that managerial interventions only produce higher levels of relative profitability in the first half year and continue to perform below the reference level for the rest of the simulation.
The neutral baseline scenario which neither changes monitoring nor type or intensity of implemented incentive schemes continues to outperform the adaptive scenarios by more than $59\%$.
The equal distribution of higher order values in the simulated workforce is reflected in the employed neutral scenario which does not favour or adversely affect the behaviour of any particular group.
This finding suggests that it might be better to not change the implemented management strategy at all and instead rely on the realisation of self-organisational capacity in the social network of the firm's employees.
Given the knowledge of the distribution of higher order values among the employees, the firm's management could anticipate which strategy would provide the best long-term fit to the emerging corporate culture that manifests itself in the employees' behaviour guided by personal values and social norms.

\section{Discussion}\label{sec: discussion}

We view the firm as a complex adaptive system, i.e. a system of a large number of agents that interact and adapt \citep{Holland.2006}. 
In our model of the firm, several kinds of interactions and adaptations occur. 
Employees interact by forming a network and by cooperating. 
Through their actions, social norms of behaviour emerge, to which each worker adapts in line with his or her own values. 
The evolving social norms anchor employees’ behaviours, but do not determine them completely. 
The management also interacts with the employees. 
The management tries to keep shirking under control, to achieve high output and to promote cooperation among employees. 
It uses direct control instruments and the parameters of a monetary reward scheme to influence employees’ actions. 
Employees indeed adapt to the management’s policies.
They change their shirking and cooperation behaviour, and the satisfaction, which influences their productivity, adapts, too. 
Finally, the management adapts its management strategy to the observed outcomes of the use of the management instruments. 
Hence, corporate culture in the form of employees’ social norms, that guide their behaviour, and management strategies (or the firm’s formal institutions) co-evolve. 
Due to the adaptation of the firm’s corporate culture, it is difficult for the management to influence the behaviour of employees in the desired way. 
Hence, there are strong constraints on the ability of the management to control the system with the given management tools.

Our analysis focused on what we call management scenarios. 
The four scenarios considered how often and how strongly the management updates the intensity and the frequency of its instrument use (or strategy). 
Our first result is that in the long run (i.e. after about six years), all scenarios converge to similar level of profitability and almost identical strategies. 
In the long run, the management does not monitor employees’ shirking anymore and uses group performance rewards as an incentive scheme. 
Over time there is an implicit learning effect that stems from the management's observations of actual employee behaviour, gradually leading to adaptation of expectations regarding shirking behaviour and produced output.
The finding that the management completely abandons monitoring efforts in the long run can therefore be interpreted as the endogenous emergence of trust \citep[comparable to the findings of][]{VandenSteen.2010a}.

Despite this convergence, there are enormous differences in profitability across the scenarios during the adaptation process.
Especially the two extreme adaptation styles – daily updating and yearly updating – lead to stark temporary differences in management strategies. 
While daily updating leads to a gradual reduction of monitoring and a relatively early increase in group rewards, yearly updating for some years generates almost the opposite strategy, i.e. high monitoring and low group rewards. 
Nevertheless, the cumulated profitability of these extreme adaptation styles are close together, indicating that both management strategies are relatively successful. 
The cumulative profitability of biannual adaptation, but especially of monthly adaptation, is clearly lower. 

The scenarios differ with regard to the impact on the different types of employees. 
Daily adaptation\footnote{
    Daily updating is more a theoretical limiting case than a realistic description of management behaviour.
    In the presence of decision-making costs, the management cannot be expected to change its strategy on a daily basis.
} leads to an early reduction of monitoring and is strongly appreciated by O-agents who in turn experience high levels of job satisfaction. 
As shown in \cite{Roos.2022}, O-agents are important for the evolution of corporate culture because their motivation or demotivation can impact others through their wide-reaching influence on social norms.
In contrast to that, the yearly updating scenario produces temporarily high individual monetary rewards which have a motivating effect on SE-agents.
In the long run, both C- and SE-agents end up with very low levels of job satisfaction because the instruments they value (i.e. monitoring for C-agents and individual rewards for SE-agents) are not used.
The mirror image of this result is that O-agents and ST-agents converge to very high levels of job satisfaction.
Because of the link between job satisfaction and labour productivity, the long-run demotivation of some employees is a crucial issue. 
In our model calibration, a long-run job satisfaction close to zero of C-agents and SE-agents implies that both groups only work with the minimum productivity of 0.5 resulting in a substantial output loss of the firm. 

The findings regarding job satisfaction require some qualifications. 
It seems plausible that employees with very low job satisfaction will quit their job at some point in time whereas there is no turnover of the firm’s workforce in our current model. 
Both the dynamics and the long-run outcomes might be quite different if employees were allowed to leave the firm when their job satisfaction falls below a threshold for a certain time. 
The different adaptation styles of the management strategy might then lead to a selection of particular types of employees in a firm. 
A firm with daily updating might quickly lose all C-agents and over time also most SE-agents. 
Vice versa, every more long-term updating would probably drive away O-agents rather soon. 
With a change of the workforce composition, we might expect that the four adaptation scenarios will not converge to the same management strategies in the long run. 
We leave a detailed analysis of workforce turnover to future work.

Another interesting finding concerns the dynamics of the interaction homophily in the endogenous social network. 
There are practically no differences across the adaptation scenarios, suggesting that the network dynamics are unaffected by the management style. 
This finding can be interpreted as a form of self-organisation that constrains the management’s ability to control the system. 
While the interaction homophily of C-, SE- and O-agents convergences to the same level, the long run value of ST-agents’ interaction homophily is twice as high. 
This means that this group has a much stronger in-groups connectivity than the others. 
ST-agents hence strongly interact among themselves implying that they develop their subculture within the firm over time. 
This result might be relevant if the management tried to influence corporate culture directly, e.g. by communication, which is not represented in our model. 
An analysis of direct efforts of the management to affect corporate culture is another interesting topic for future research.

A final remarkable result is that all adaptation scenarios with changing management styles lead to significantly lower cumulated profitability than a baseline scenario in which the management initially chooses a neutral management style and sticks with it forever. 
Neutral management style means that the management chooses an intermediate monitoring strategy and abstains from using pay-for-performance schemes. 
The key point is that while it is possible to influence employees’ behaviour with monetary incentives, using rewards is also costly. 
Furthermore, a constant strategy makes it easier for social norms to converge quickly on a dominant path \citep[see][]{Roos.2022}. 
Hence letting the self-organisation forces within a firm work might be preferable to active management efforts to achieve a certain behaviour. 
This is in line with the considerations of \citet[page 167]{Li.2021a} that \enquote{culture is not just one of the inputs to strategy or part of its execution} but that it \enquote{competes with strategy for control} and as such they can function as direct substitutes.
However, we consider this conclusion as tentative and preliminary and stress that more research is necessary to check its robustness to variations of the model.

In future work, our model could be modified in several dimensions. 
First, employees only differ in terms of their values, but not in terms of other characteristics such as skills or knowledge. 
As a consequence, the task interdependence and hence the necessary cooperation is rather abstract. 
In the present model, it is not necessary that certain employees cooperate due to skill or knowledge complementarities, which is a limitation. 
Relatedly, there is no hierarchy and no formal working structure in the firm whereas both might have an impact on the formation of norms and the evolution of intra-firm subcultures.
Second, there is no labour turnover.
Employees do not quit if they are dissatisfied and the management does not fire underperformers or employees who received several written warnings as a result of being caught shirking more than deemed acceptable.
When employees are allowed to leave the firm, a hiring process of new employees must also be modelled. 
By selecting employees according to their value types, the management would have another management instrument that might be crucial for performance. 
Third, the management is modelled as an abstract entity, but in reality it consists of individuals with values and behaviours as well. 
Modelling managers as individuals could also have a direct impact on corporate culture, either by being a role model or through direct communication efforts. 
Finally, in the current model, both monitoring and the updating of the management strategy are practically costless.
However, monitoring efforts would cause a direct resource cost impacting profitability while changing the management strategy would require efforts from the managers and entail learning and implementation costs.
The latter is already partially incorporated in the model as visible in the adaptation speed (or stickiness) of social norms.
Additionally, these aspects might also depend on personal characteristics of the managers and on power relations within the firm. 

\section{Conclusion}\label{sec: conclusion}

Our paper shows that a firm can be viewed and modelled as a complex adaptive system. 
Due to the adaptation of employees to the management strategy, the emergence of social norms and self-organisation within the workforce, the management’s ability to control the firm is limited. 
We define corporate culture as the endogenous social norms that regulate employees’ shirking and cooperation behaviour, which in turn has a direct impact on the firm’s output. 
Employees’ responses to social norms are driven by their values. 
The management tries to influence employees’ behaviour directly through monitoring and monetary incentives, but does not consider the indirect effects on corporate culture. 
This implies that management policies can have unintended side effects which counteract the direct ones.
The presented model provides plenty of opportunities for future extensions, e.g. by adding personal skills and knowledge, a formal hierarchy and dependencies in the organisation, costs to monitoring and strategy changes, or fluctuations in the workforce.

We show that firms with a management that adopts extreme adaptation styles of their management strategy have higher profitability than firms that chose intermediate or moderate adaptation styles.
Firms in which adaptation occurs either very frequently (i.e. daily) or very infrequently (i.e. yearly) have higher cumulated profitability at the end of the simulations.
The different adaptation styles have diverse effects on employees with different values and hence on endogenous corporate culture. 
We find that adaptation of the management style leads to a long-run decrease in monitoring and the increased use of group performance rewards. 
The decrease in monitoring can be interpreted as the endogenous emergence of trust of the management in employees. 
Frequent adaptation with a fast decrease of monitoring has a strong positive effect on the satisfaction and the performance of employees who are self-directed and open to change. 
Due to their connectedness inside the firm's social network, they are drivers of corporate culture. 
However, we also find that active adaptation of the management’s strategy is always inferior with regard to a firm's profitability than a non-adapting management style that is already tailored to fit the value composition in the workforce. 
In firms with non-adapting management the self-organisation of corporate culture and its effect on employee behaviour is not disturbed.

% ----- end main body

\bibliographystyle{chicago}
\bibliography{FirmAsCAS}

% ----- begin appendix

\newpage

\begin{appendices}
\counterwithin{figure}{section}
\counterwithin{table}{section}

\setcounter{equation}{0} % temporary

\section{Additional tables} \label{sec: app-1}
\setcounter{table}{0}
% This table should discriminate between the three modules
\begin{table}[H]
    \centering
    \resizebox{!}{.2\textwidth}{\begin{tabular}{llccc}
    \toprule
    Parameter & Description & \multicolumn{3}{c}{Value}\\ \cmidrule{3-5}
    & & Social Network & Adaptive Behaviour & Endogenous Management\\
    & & (Section \ref{subsec: network}) & (Section \ref{subsec: adaptive-behaviour}) & (Section \ref{subsec: end_management})\\
    \hline \hline
        $n$ & number of agents in the model & $100$ & — &— \\
         $dist$ & distribution of agent types & $(0.25, 0.25, 0.25, 0.25)$& —&— \\
    $\kappa$ & degree of task interdependence & $0.5$& —&—\\
       $w_{b}$ & hourly wage & $1$& —&— \\
         $h$ & rate of adjustment to social norms & $0.1$&— &— \\
         $\tau$ & maximum available time & $8$& —&— \\
        % End. Network + Adaptive
      $\Sigma$ & monitoring intensity & $\{0, 0.5, 1\}$& — & $\Sigma \rightarrow \Sigma_{t} \in [0,1]$ \\
       $\mu$& incentive intensity & $\{0, 1\}$& — & $\mu \rightarrow \mu_{t} \in [0,1]$ \\
       $\lambda$ & type of PFP scheme & $\{0, 0.5, 1\}$& — & $\lambda \rightarrow \lambda_{t} \in [0,1] $\\
      % Adaptive + End. Manag.
      $S^{eff}$ & mediator satisfaction &/ & $0.5$& —\\
      $ETC$& subset of observed employees &/& $ETC \subset N$& $ETC \rightarrow ETC_{t}$\\
      $\eta$ & shock of being caught &/& $0.05$& — \\
      % End. Manag.
      $suf$ & strategy update frequency &/&/& $\{1, 30, 180, 365\}$\\
      $sui$ & strategy update intensity &/&/& $\{\frac{1}{600}, \frac{1}{20}, \frac{3}{10}, \frac{73}{120}\}$\\

    \bottomrule
    \end{tabular}}
     \caption{\enspace Model parameters. We use "/"  to indicate a missing element, "—" no changes from the previous column, and "$\rightarrow$" for endogenised parameters.}
    \label{tab: parameters}
\end{table}

\begin{landscape}
 \begin{table}[p!]
   \begin{adjustbox}{width=1.5\textwidth}
   \begin{tabular}{lcccc}
 \toprule
     Description   & Original Equations & \multicolumn{3}{c}{Model Equations} \\ \cmidrule{3-5}
         &\citep{Roos.2022} & Social Network& Adaptive Behaviour  & Endogenous Management \\
         & &  (Section \ref{subsec: network}) & (Section \ref{subsec: adaptive-behaviour}) & (Section \ref{subsec: end_management}) \\\hline \hline
 1. \hyperlink{page.12}{Individual output} & $O_{i,t} = p_{i,t}^{(1-\kappa})*\bar{c_{i,t}}^{\kappa}$ & — &  $O_{i,t} = \pi_{i,t} (p_{i,t}^{\phantom{i}(1-\kappa)} * \bar{c}_{i,t}^{\phantom{i}\kappa})$& —\\

 2. \hyperlink{page.12}{Productivity effects} &/ & / & $\pi_{i,t} = (1 - S^{eff}) + 2 \cdot S^{eff} \cdot S_{i,t}$& — \\
 3. Average cooperation & $ \bar{c}_{i,t}= \sfrac{1}{(n-1)} \sum_{j \neq i} c_{j,t}$ & — & — & — \\

 4. \hyperlink{page.11}{Shirking time norm} & $ s^{*}_{t} = (1-h)\; s^{*}_{t-1} + h\; \frac{\sum_{j \epsilon N} s_{j,t-1}}{n}$ & $ s_{i,t}^{*} =
    \left\{ \begin{array}{ll}
        (1 - h) \; s_{i,t-1}^{*} + h \; \frac{\sum_{j \in I_{i,t}} \Delta e_{i,j,t-1} \; s_{j,t-1}}{\sum_{j \in I_{i,t}} \Delta e_{i,j,t-1}} & \text{if} \; I_{i,t} \neq \emptyset\\
        s_{i,t-1}^{*}      & \text{otherwise}
    \end{array} \right.$& — & —\\

 5. \hyperlink{page.11}{Cooperative time norm} & $ c^{*}_{t} = (1-h)\; c^{*}_{t-1} + h\; \frac{\sum_{j \epsilon N} c_{j,t-1}}{n} $ & $c_{i,t}^{*}  =
    \left\{ \begin{array}{ll}
        (1 - h) \; c_{i,t-1}^{*} + h \; \frac{\sum_{j \in I_{i,t}} \Delta e_{i,j,t-1} \; c_{j,t-1}}{\sum_{j \in I_{i,t}} \Delta e_{i,j,t-1}} & \text{if} \; I_{i,t} \neq \emptyset\\
        c_{i,t-1}^{*}      & \text{otherwise}
    \end{array} \right.$ & — & — \\

6. Individual time norm & $p_{t}^{*} = \tau - s_{t}^{*} - c_{t}^{*}$ & $p_{i,t}^{*} = \tau - s_{i,t}^{*} - c_{i, t}^{*}$ & — & — \\

7. Shirking time & $s_{i,t} \sim T(a_{i,t}, b_{i,t}, m_{i,t})$ & — & — & —\\

8. Cooperative time & $c_{i,t} \sim T(a_{i,t}, b_{i,t}, m_{i,t})$ & — & — & —\\

9. Individual time & $p_{i,t} = \tau - s_{i,t} - c_{i,t}$ & — & — & —\\

10. Lower bound triangular distribution & $a_{i,t} = x_{t}^{*}(1-\delta_{i})$, where $x_{t}^{*} = \{s_{t}^{*}, c_{t}^{*}\}$ & — & — & —\\

11. \hyperlink{page.14}{Upper bound triangular distribution} & $b_{i,t} = x_{t}^{*}(1+\delta_{i})$ , where $x_{t}^{*} = \{s_{t}^{*}, c_{t}^{*}\}$& — & $b_{i,t} = x_{i,t}^{*}(1+\beta_{i,t\phantom{i}}\delta_{i})$, where $x_{i,t}^{*} = \{s_{i,t}^{*}, c_{i,t}^{*}\}$ & —\\

12. Mode triangular distribution & $ m_{i,t} = \left \{\begin{array}{ll} x^{*}_{t} + \phi_{i,t} & \text{if}\; x_{t}^{*} = s_{i,t}^{*}\\
x^{*}_{t} + \gamma_{i,t} + \rho_{i,t} & \text{if}\; x_{t}^{*} = c_{i,t}^{*}  \\
\end{array}\right.$, where $x_{t}^{*} = \{s_{t}^{*}, c_{t}^{*}\}$ &
$x_{t}^{*} \rightarrow x_{i,t}^{*}$
& — & —\\

13. Deviation from norms & $ \delta_{i} = \left \{\begin{array}{ll}
\sfrac{1}{3}   & \text{if}\; i = C  \\
1   & \text{if} \; i = O  \\
\sfrac{2}{3}   & \text{if}\; i = SE  \\
\sfrac{2}{3}   & \text{if}\; i = ST  \\
\end{array}\right. $ & — & — & —\\

14. Need for autonomy & $ \phi_{i,t} = \left \{\begin{array}{ll}
\phantom{-}0.5s_{t}^{*}\delta_{i} & \text{if}\; i = C \; \text{and}\; \Sigma_{t} = 0\; \text{or}\; i = O \; \text{and}\; \Sigma_{t}= 1\\
-0.5s_{t}^{*}\delta_{i}   & \text{if}\; i = C \; \text{and}\; \Sigma_{t} = 1\; \text{or}\; i = O \;\text{and} \; \Sigma_{t} = 0\\
\phantom{-}0 & \text{otherwise} \\
\end{array}\right.$
    &
$s_{t}^{*} \rightarrow s_{i,t}^{*}$
& —
& $\Sigma = 0 \rightarrow \Sigma_{t} \in \left[0,0.5\right)$ and $\Sigma = 1 \rightarrow \Sigma_{t} \in \left(0.5,1\right]$\\

15. Degree of cooperativeness & $ \gamma_{i,t} =  \left \{\begin{array}{ll}
-0.5c_{t}^{*}\delta_{i}   & \text{if}\; i = SE  \\
 \phantom{-}0.5c_{t}^{*}\delta_{i}   & \text{if} \; i = ST  \\
 \phantom{-}0 & \text{otherwise} \\
\end{array}\right.$ &
$ c_{t}^{*} \rightarrow c_{i,t}^{*}$ & —& —\\

16. Responsiveness to rewards & $ \rho_{i,t} = \left \{\begin{array}{ll}
-0.5c_{t}^{*}\delta_{i}   & \text{if}\; i = SE \; \text{and}\; \lambda = 0\\
-0.1c_{t}^{*}\delta_{i}   & \text{if}\; i = ST \; \text{and}\; \lambda = 0\\
 \phantom{-}0.1c_{t}^{*}\delta_{i}   & \text{if}\; i = SE \; \text{and}\; \lambda = 1\\
\phantom{-}0.5c_{t}^{*}\delta_{i}   & \text{if}\; i = ST \; \text{and}\; \lambda = 1\\
\phantom{-}0 & \text{otherwise} \\
\end{array}\right.$
    &
$c_{t}^{*} \rightarrow c_{i,t}^{*}$ & — &
$\lambda = 0 \rightarrow \Sigma_{t} \in \left[0,0.5\right)$ and $\lambda = 1 \rightarrow \Sigma_{t} \in \left(0.5,1\right]$\\

17. \hyperlink{page.12}{Employees' bonus} & $B_{i,t} = (1-\lambda)O_{i,t} + \lambda (\frac{1}{n}) \sum_{j=1}^{n} O_{j,t}$ & — & — & $\lambda \rightarrow \lambda_{t}$\\

18. \hyperlink{page.12}{Financial rewards} & $ R_{i,t} = \omega_{b} + \mu B_{i,t}$ & — & — & $ \mu \rightarrow \mu_{t}$\\
        \bottomrule
\end{tabular}
\end{adjustbox}
     \caption{Model equations overview. Comparison between the original equations in \cite{Roos.2022} and the three additional extensions. We use "/" to indicate a missing element, "—" no changes from the equation in the column to the left, and "$\rightarrow$" for changes of single variables. \textit{Continues on the next page.}}
    \label{tab: list_eq}
\end{table}
\end{landscape}

\begin{landscape}
 \begin{table}[p!]
\begin{adjustbox}{width=1.5\textwidth}
   \begin{tabular}{lcccc}
 \toprule
     Description   & Original Equations & \multicolumn{3}{c}{Model Equations} \\ \cmidrule{3-5}
        &\citep{Roos.2022} & Social Network& Adaptive Behaviour  & Endogenous Management \\
         & &  (Section \ref{subsec: network}) & (Section \ref{subsec: adaptive-behaviour}) & (Section \ref{subsec: end_management}) \\\hline \hline

 19. \hyperlink{page.16}{ Management style} & $\Sigma = \{0,0.5,1\}$& — & — & $   \Sigma_{t} =
    \left\{ \begin{array}{ll}
    (1 + sui) \Sigma_{t-1} & \text{if}\; \frac{1}{x}\sum\limits_{t-x}^{t-1} \bar s_{t} > s_{t}^{max}\\
    (1 - sui) \Sigma_{t-1} & \text{if}\; \frac{1}{x}\sum\limits_{t-x}^{t-1} \bar s_{t} \leq s_{t}^{max}\\
    (1 - \frac{\bar O_{t}}{EGO_{t}}) sui  & \text{if}\; \Sigma_{t-1} = 0 \; \wedge \; \frac{1}{x}\sum\limits_{t-x}^{t-1} \bar O_{t} < EGO_{t}
    \end{array}
    \right.$\\

20. \hyperlink{page.17}{PFP schemes} & $\lambda = \{0,0.5,1\}$ &— & — & $   \lambda_{t} =
    \left \{ \begin{array}{ll}
    (1 + sui) \lambda_{t-1} &\text{if}\; \frac{1}{x}\sum\limits_{t-x}^{t-1} \bar c_{t} < \kappa \cdot \alpha_{t} \\
    (1 - sui) \lambda_{t-1} &\text{if}\; \frac{1}{x}\sum\limits_{t-x}^{t-1} \bar c_{t} > \kappa \cdot \alpha_{t} \\
    \lambda_{t-1}  & \text{otherwise} \\
    \end{array} \right.$\\

21. \hyperlink{page.17}{Intensity of rewards} & $\mu = \{0,1\}$ & — & — & $\mu_{t} =
    \left\{ \begin{array}{ll}
    (1 + sui) \mu_{t-1} & \text{if}\; \frac{1}{x}\sum\limits_{t-x}^{t-1} \bar O_{t} < EGO_{t} \\
    (1 - sui) \mu_{t-1} & \text{if}\; \frac{1}{x}\sum\limits_{t-x}^{t-1} \bar O_{t} > EGO_{t} \\
    \mu_{t-1} & \text{otherwise} \\
    \end{array}\right.$\\

22. \hyperlink{page.8}{Weighted absolute-deviation index (WADI)} & / & $AD_{i,j,t} = \frac{\sum_{k=1}^{A} \lvert a_{k,i,t} - a_{k,j,t} \rvert}{\tau}$ & — & —\\

23. \hyperlink{page.9}{Activity similarity} & / & $ AS_{i,j,t} = 1 - AD_{i,j,t}$ & — & —\\

24. \hyperlink{page.9}{Chance of interaction} & / & $d_{t} = \{ d_{i,j,t} \sim U(0, 1), \; \forall (i,j) \in N \}$ & — & — \\

25. \hyperlink{page.10}{Potential interactions check} & / & $
    I_{i,t}^{pot} =
      \{j \;|\; e_{i,j,t-1} > e_{i,k,t-1}, \forall \; j,k \in P_{i,t-1},\; j\neq k \} \cup
       \{j \; | \; j \in_{R} N \setminus P_{i,t-1} \}$ & — & —\\

26. \hyperlink{page.10}{Set interacting agents} & / & $ I_{i,t} = \{ j \; | \; d_{i,j,t} \; < \; AS_{i,j,t}, j \neq i \}$ & — & —\\

27. \hyperlink{page.10}{Weight change agents' directed edge} & / & $\Delta e_{i,j,t} = \left\{ \begin{array}{ll}
AS_{i,j,t} & \text{if} \; j \in I_{i,t}\\
0          & \text{otherwise}
\end{array} \right.$ & — & — \\

28. \hyperlink{page.11}{Network connections} & / & $  e_{i,j,t} =
    \left\{ \begin{array}{ll}
        \frac{(t-1) \cdot e_{i,j,t-1} + \Delta e_{i,j,t}}{t} & \text{if}\; t \geq 1 \\
        0                                                    & \text{if}\; t = 0
    \end{array} \right.$ & — & — \\

29. \hyperlink{page.13}{Base satisfaction levels} & / & / & $   S_{i}^{0} =
      S_{i}^{0} =
    \left\{ \begin{array}{ll}
        \Sigma                    & \text{if}\; i\;\in\; \text{C-type} \\
        1 - \Sigma                & \text{if}\; i\;\in\; \text{O-type} \\
        0.5 + \mu (0.5 - \lambda) & \text{if}\; i\;\in\; \text{SE-type} \\
        0.5 + \mu (\lambda - 0.5) & \text{if}\; i\;\in\; \text{ST-type}
    \end{array} \right.$ & $\{\Sigma, \lambda, \mu \} \rightarrow \{\Sigma_{t}, \lambda_{t}, \mu_{t} \}$\\

30. \hyperlink{page.13}{Satisfaction adjustments} & / & / & $ S_{i,t} =
    \left\{ \begin{array}{ll}
        0.99 \; S_{i,t-1} & \text{if}\; S_{i,t-1} > S_{i}^{0} \\
        1.01 \; S_{i,t-1} & \text{if}\; S_{i,t-1} < S_{i}^{0}
    \end{array} \right.$ & —\\

31. \hyperlink{page.15}{Written warning scaling factor} & / & / & $  \beta_{i,t}=
    \left\{ \begin{array}{lll}
    1 - \frac{|WW_{i,t}|}{3} + \frac{|WW_{i,t}|}{3} \; \frac{t - x_{n}}{t} & \text{if}\; 0 \leq |WW_{i,t}| < 3\\
    0.0 + 1.0 \; \frac{t - x_{n}}{t} & \text{otherwise}\\
    \end{array} \right.$ & —\\

32. \hyperlink{page.16}{Maximum acceptable shirking} & / & / & $s^{max} = \sfrac{\tau}{10}$ & $ s_{t}^{max} = (1 - h) \; s_{t-1}^{max} + h \; \frac{1}{n} \sum_{i}^{ETC_{t}} s_{i,t}$\\

33. \hyperlink{page.15}{Observed shirking time} & / & / & / & $ \bar s_{t} = \frac{1}{n} \sum_{i}^{ETC_{t}} s_{i,t}$\\

34. \hyperlink{page.15}{Observed cooperative time} & / & / &/ & $ \bar c_{t} = \frac{1}{n} \sum_{i}^{ETC_{t}} c_{i,t}$\\

35. \hyperlink{page.16}{Expected group output} & / & / & / &  $EGO = \left[\alpha_{t}(1-\kappa)\right]^{(1-\kappa)} \cdot (\alpha_{t}\kappa)^{\kappa}$\\
    \bottomrule
    \end{tabular}
\end{adjustbox}
    \caption*{Table \ref{tab: list_eq} Continued: We use "/" to indicate a missing element, "—" no changes from the equation in the column to the left, and "$\rightarrow$" for changes of single variables.}
\end{table}
\end{landscape}

\newpage
\section{Additional figures} \label{sec: app-2}

\begin{figure}[H]
\centering
\includegraphics[scale = 0.35]{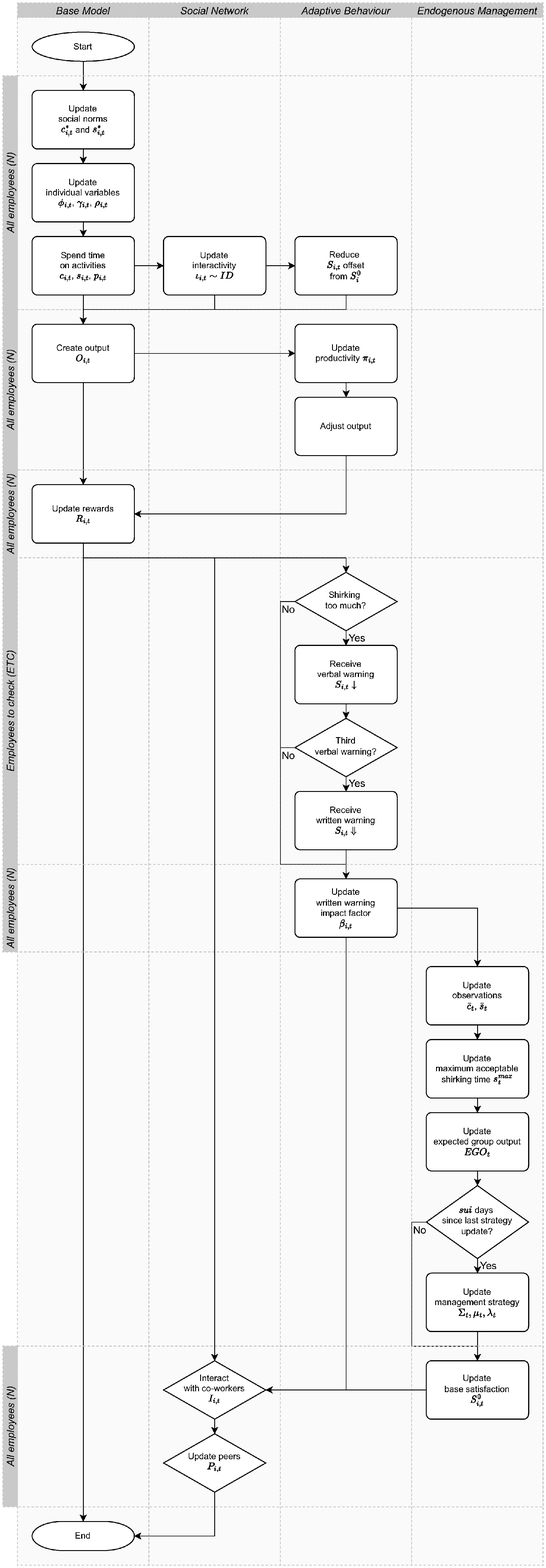}
\caption{Model stepping flowchart. Source: Authors' own illustration.}
\label{fig: flowchart}
\end{figure}

\newpage
\section{Sensitivity Analyses} \label{sec: app-3}

Initial values for model parameters and agent variables are important in complex adaptive systems because they can influence the outcome of the computational simulations through pre-determined path dependencies.
To account for this, we conducted sensitivity analyses on the three management strategy parameters (Appendix \ref{subsec: app-3-strategies}) as well as on the method of initial time allocation used for all employees (Appendix \ref{subsec: app-3-time-init-method}).
All parameter scans define starting values for the model initialisation, i.e. they determine on the initial step of the simulation.
The simulations have been run 100 times for each unique parameter combination and the mean values across those replicates are presented in the plots below.

\subsection{Initial management strategies}\label{subsec: app-3-strategies}
\counterwithin{figure}{subsection}

The parameters $\Sigma$ (monitoring), $\mu$ (PFP intensity), and $\lambda$ (PFP type) have been altered one-by-one to isolate them from one another and paint a clearer picture regarding their individual effects on the firm's profitability.
Each parameter has been checked for the values $0.0, 0.25, 0.5, 0.75, 1.0$.

\begin{figure}[H]
\centering
\includegraphics[width = 0.75\textwidth]{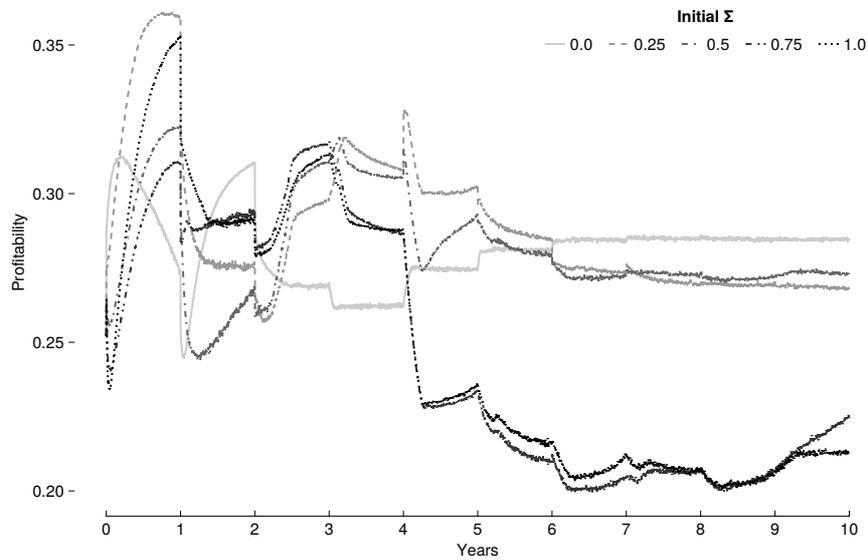}
\caption{Firm profitability over time for five initial values of monitoring strategy $\Sigma$. Source: Authors' own illustration.}
\label{fig: sigma-profitability}
\end{figure}

\begin{figure}[H]
\centering
\includegraphics[width = \textwidth]{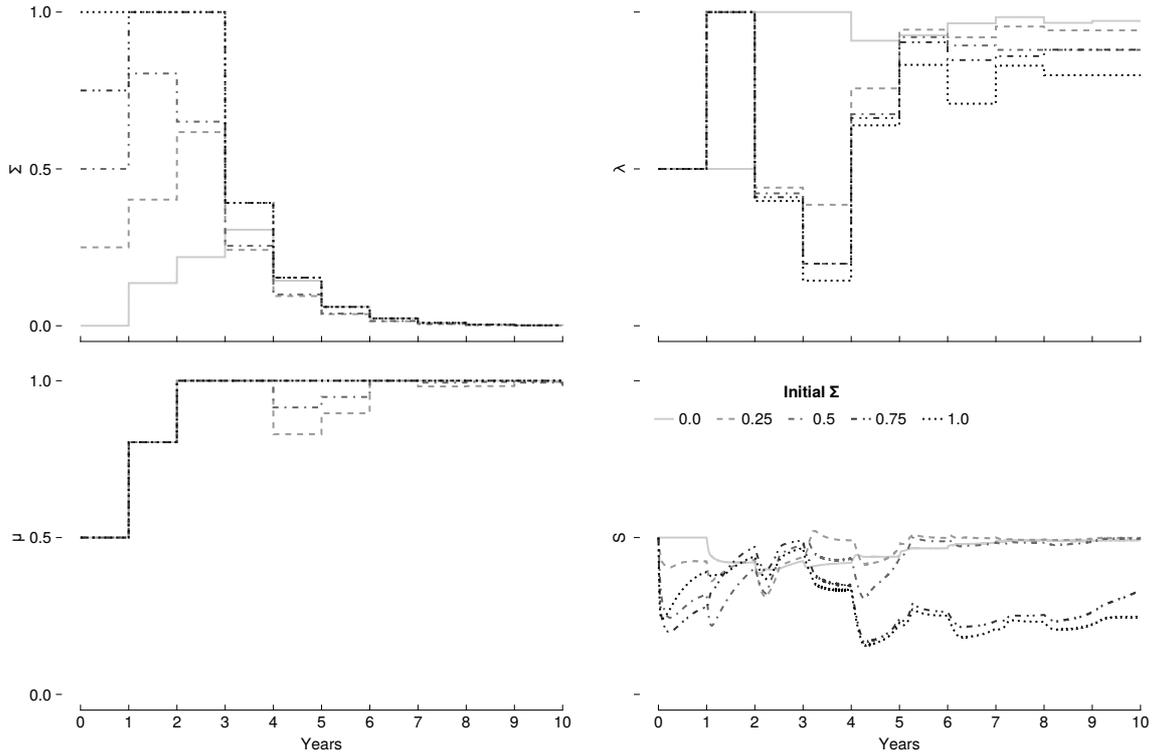}
\caption{Monitoring ($\Sigma$), PFP intensity ($\mu$), PFP style ($\lambda$), and mean employee satisfaction ($S$) over time for five initial values of monitoring strategy $\Sigma$. Source: Authors' own illustration.}
\label{fig: sigma-combined}
\end{figure}

Figure \ref{fig: sigma-profitability} reveals that the first years of the simulation show similar developments for all $5$ tested initial $\Sigma$ values.
At the beginning of year $4$, there is a secession of the lines' trajectories, separating the results into two groups: initial $\Sigma$ above and below/equal $0.5$.
This grouping happens due to two separate reactions from the workforce to management style changes as well as a path dependency due to our modelling decisions (see Figure \ref{fig: sigma-combined}).

First, PFP intensity $\mu$ is reduced from $1.0$ by about $10$-$20$ \%P, leading to lower individual rewards based on the implemented PFP scheme, and thus also a better profitability ratio.
Lower controlling measures at the start of a simulation allow the self-organising workforce to form social norms which after a few years are so beneficial in terms of firm output, that the management's expectations are fulfilled.
Hence, it doesn't see the need to incentivise specific behaviour according to the current type of PFP scheme as strongly anymore and a reduction of $\mu$ is issued.
The following drop in base satisfaction $S_i^0$ of SE-type agents, who generally spend their time productively due to their embeddedness in the firm's workforce, significantly reduces firm output below the threshold desired by the management and $\mu$ is again gradually increased over the subsequent years.

Second, PFP type $\lambda$ is generally higher for those simulations with lower initial $\Sigma$ after year $2$.
This suggests that initially less controlling environments allow self-organisation among employees to come into effect, evidently leading to individual behaviour (and thus descriptive social norms) leaning towards personal over cooperative tasks.
A strong upwards shift in $\lambda$ after year $4$, together with the aforementioned downward movement of $\mu$, does not yield the intended effect of increasing cooperative behaviour throughout the workforce.
This is because the ST-type agents susceptible to such PFP measures have very high levels of interaction homophily which prohibits their cooperative behaviour to spread via emergent social norms.

All initial values of $\Sigma$ gradually decline and ultimately converge towards $0.0$.
But due to how we modelled management updates (i.e. relative changes to current value), higher initial values of $\Sigma$ take longer to decline to similar levels as lower initial values.
Over time, this leads to more employees getting caught shirking too much and receiving written warnings which in turn temporarily reduce their satisfaction and consequently also their productivity.
This suggests that management's initial monitoring efforts strongly determine the outcome of our simulations through path-dependencies as well as sustained effects on worker productivity.
Further cautiously interpreting the results of this sensitivity analysis, the impression arises that in the long run it is best for a firm's productivity if its management starts off with very low monitoring efforts and by that leave the employees to self-organise into a productive work environment (cf. \cite{Roos.2022} for similar results).

\begin{figure}[H]
\centering
\includegraphics[width = 0.75\textwidth]{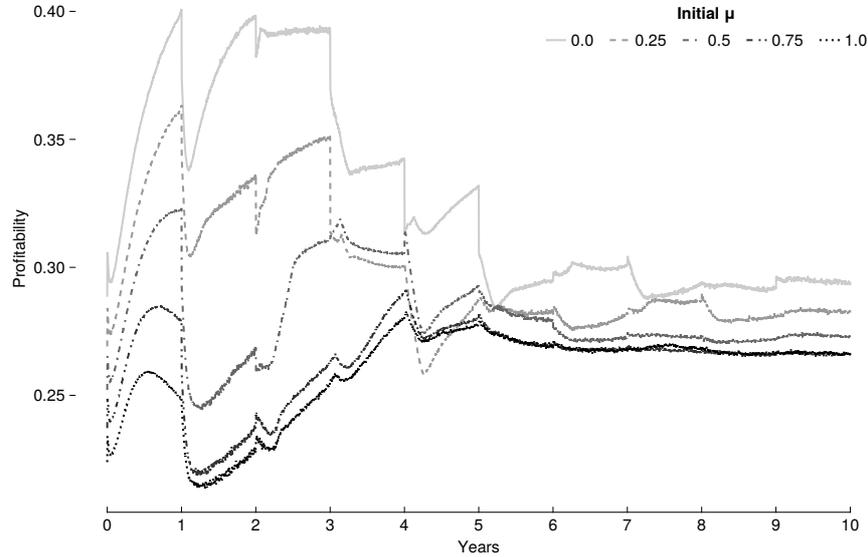}
\caption{Firm profitability over time for five initial values of PFP intensity $\mu$. Source: Authors' own illustration.}
\label{fig: mu-profitability}
\end{figure}

\begin{figure}[H]
\centering
\includegraphics[width = \textwidth]{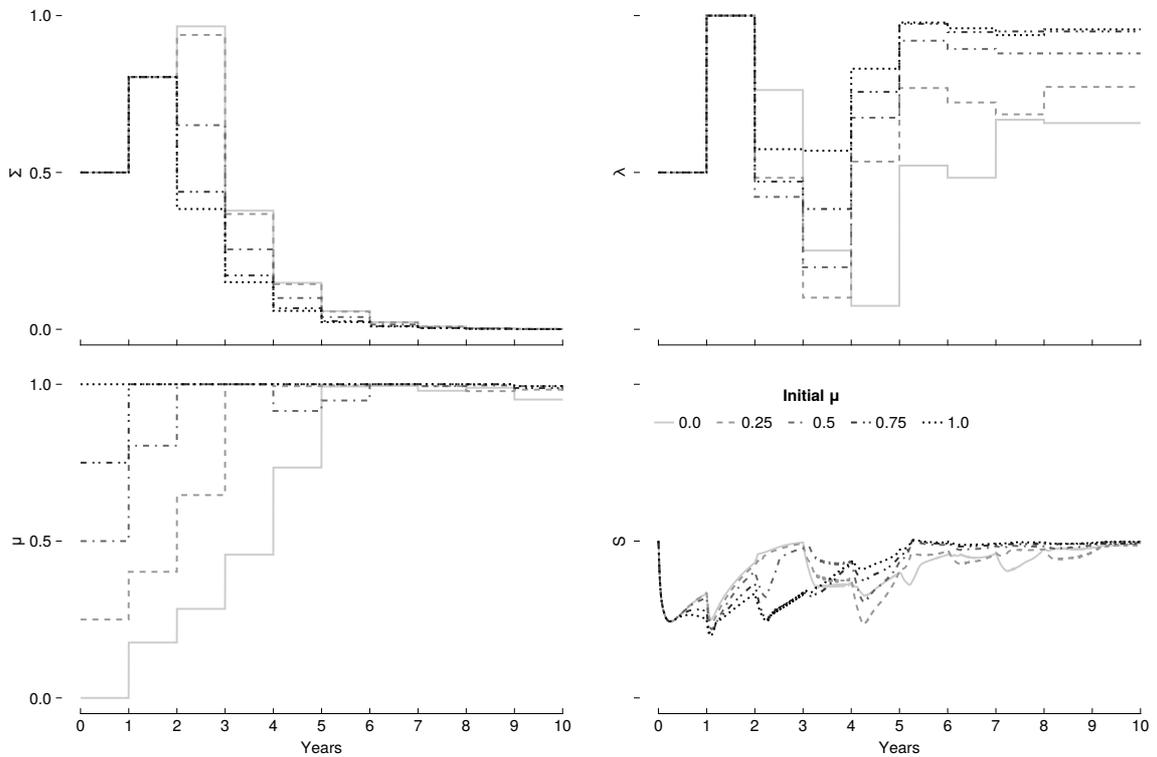}
\caption{Monitoring ($\Sigma$), PFP intensity ($\mu$), PFP style ($\lambda$), and mean employee satisfaction ($S$) over time for five initial values of PFP intensity $\mu$. Source: Authors' own illustration.}
\label{fig: mu-combined}
\end{figure}

The results in Figures \ref{fig: mu-profitability} and \ref{fig: mu-combined} show that lower initial $\mu$ leads to higher profitability in the long run.
They are in line with theoretical considerations as well as the results described by \cite{Roos.2022}.

\begin{figure}[H]
\centering
\includegraphics[width = 0.75\textwidth]{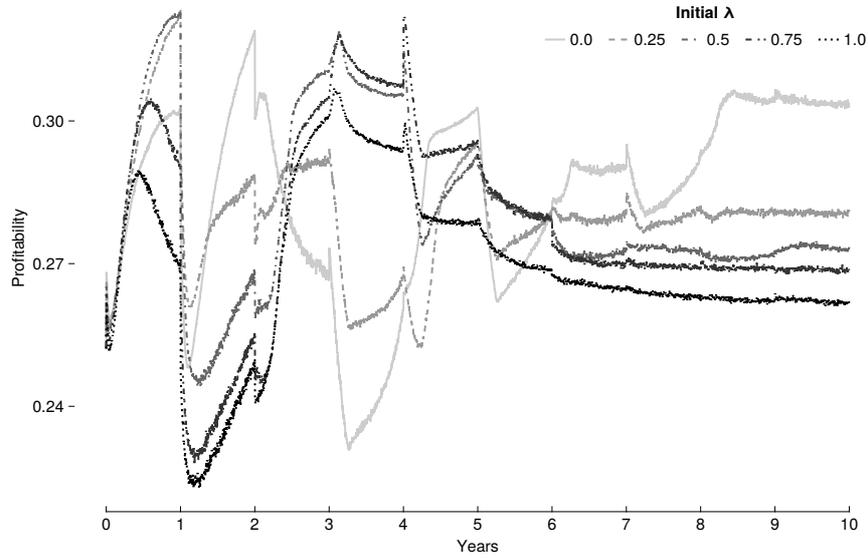}
\caption{Firm profitability over time for five initial values of PFP style $\lambda$. Source: Authors' own illustration.}
\label{fig: lambda-profitability}
\end{figure}

\begin{figure}[H]
\centering
\includegraphics[width = \textwidth]{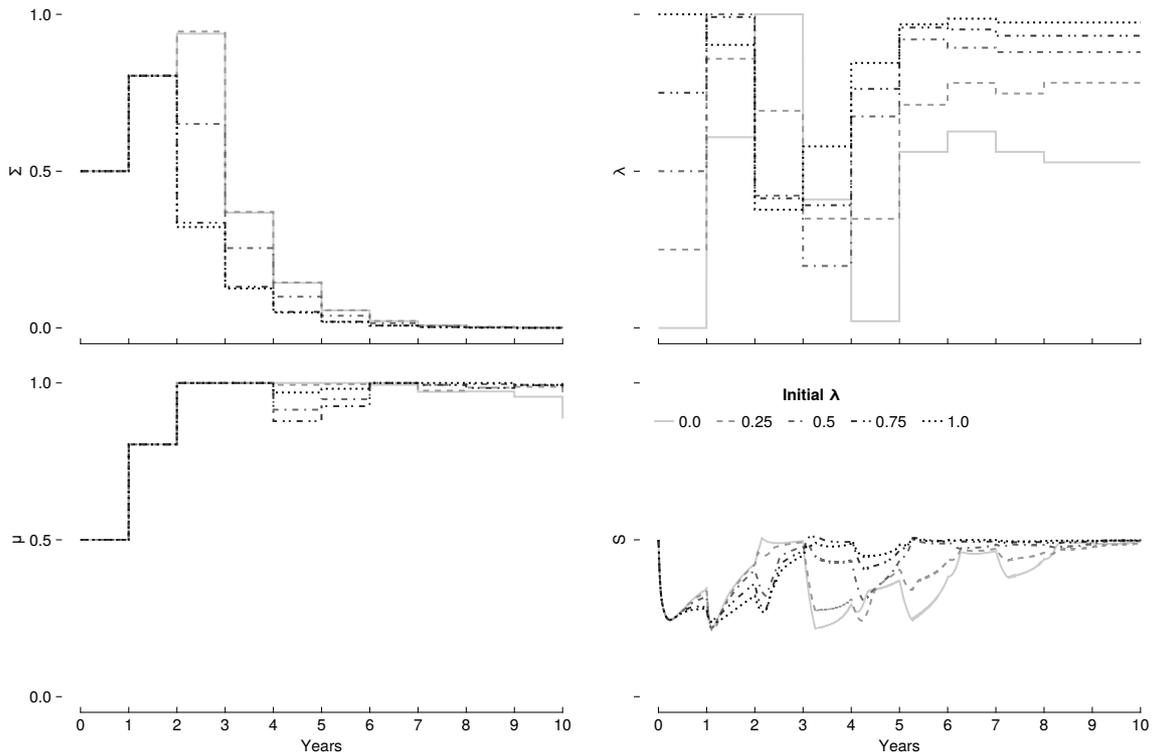}
\caption{Monitoring ($\Sigma$), PFP intensity ($\mu$), PFP style ($\lambda$), and mean employee satisfaction ($S$) over time for five initial values of PFP style $\lambda$. Source: Authors' own illustration.}
\label{fig: lambda-combined}
\end{figure}

The results in Figures \ref{fig: lambda-profitability} and \ref{fig: lambda-combined} show that lower initial $\lambda$ leads to higher profitability in the long run.
They are in line with theoretical considerations as well as the results described by \cite{Roos.2022}.

\subsection{Initial time allocation method}\label{subsec: app-3-time-init-method}

Our model assumes a "blank slate" firm for the theoretical simulations.
This is somewhat akin to a freshly incorporated firm that has hired $n$ employees who do not know each other before beginning to work there.
How agents spend their time in the first step is therefore of crucial importance for the simulation results.
The method for employees' initial time allocation determines $s_{i,t=0}$, $c_{i,t=0}$, and residually $p_{i,t=0}$.
We tested four possible initialisation methods:

\begin{enumerate}
    \item \textbf{Randomly}: Randomised initial time allocation;
    \item \textbf{Equally}: Equal initial time allocation, i.e. $\sfrac{\tau}{3}$ for each activity;
    \item \textbf{Kappa}: Initial time allocation is informed by the firm's task interdependency $\kappa$ and the maximum acceptable shirking $S_{t}^{max}$;
    \item \textbf{Kappa, no shirking}: Initial time allocation is informed solely by the firm's task interdependency $\kappa$.
\end{enumerate}

\begin{figure}[H]
\centering
\includegraphics[width = 0.7\textwidth]{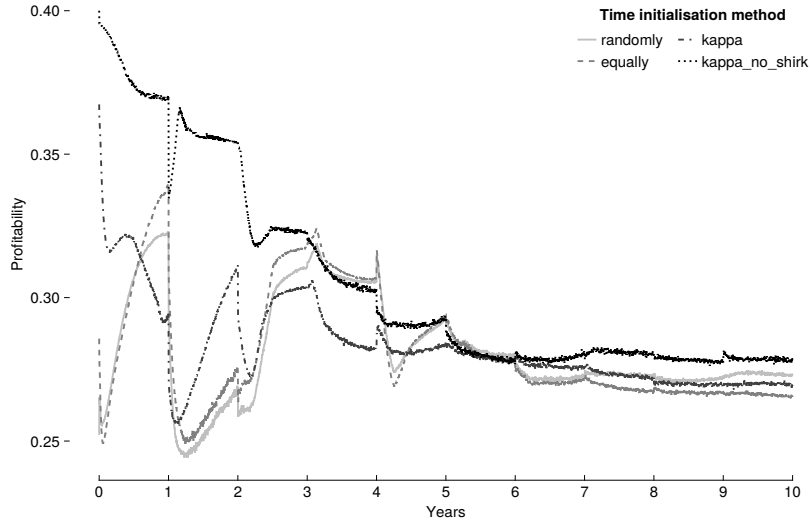}
\caption{Profitability over time for four methods of employees' initial time allocation. Source: Authors' own illustration.}
\label{fig: time-init-method-profitability}
\end{figure}

Figure \ref{fig: time-init-method-profitability} reveals that there are visible differences between the four methods at the end of the simulation, with stabilised sideways trends settling in from year 6 onwards as well as similar final profitability both in terms of level and low recent volatility.

In the main results for this paper, we have chosen the method of randomly setting employees' initial time allocations because it (i) requires the least assumptions, and (ii) still allows to eliminate stochastic effects by averaging over our replicate model runs.
Figure \ref{fig: time-init-method-profitability} displays that this version starts at the lowest profitability, is the worst performing in the short and medium run, and finally comes out on second place after 8+ years.

\begin{figure}[H]
\centering
\includegraphics[width = 0.7\textwidth]{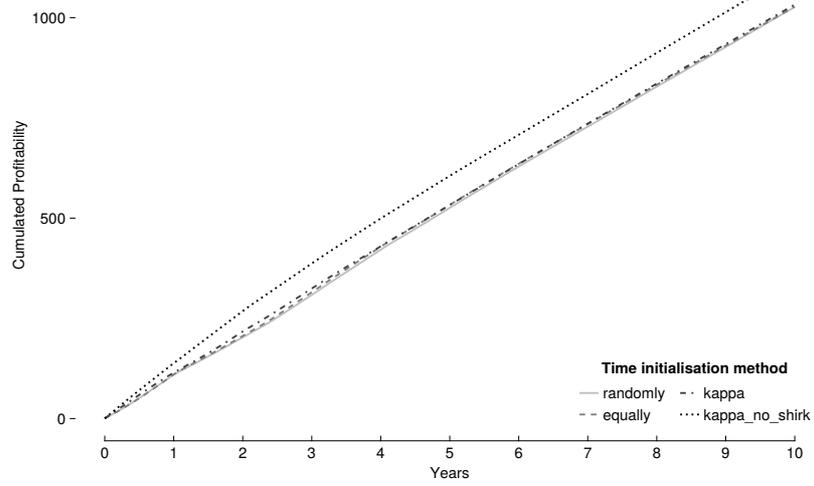}
\caption{Cumulated profitability over time for four methods of employees' initial time allocation. Source: Authors' own illustration.}
\label{fig: time-init-method-profitability-cumulated}
\end{figure}

The \textit{Kappa} and \textit{Kappa, no shirking} methods naturally start at significantly higher profitability levels compared to the other two methods.
However, their advantages dwindle after 1-3 years, effectively negating the initialisation-induced headstart (see Figure \ref{fig: time-init-method-profitability-cumulated} for the persistent effect on cumulated profitability).
Indeed, the \textit{Kappa} method exhibits the strongest but also least erratic downward trend over time which still produces the best long run outcome for firm profitability after $10$ years.
This is likely due to reduced overall shirking behaviour stemming from the absence of any shirking at step $0$ which inhibits the development of detrimental social norms on shirking.

\end{appendices}

% ----- end appendix

\end{document}